\documentclass{jaa}
\usepackage{fix-cm}
\usepackage[T1]{fontenc}
\usepackage{ae,aecompl}
\usepackage[normalem]{ulem}
\usepackage{graphicx}	
\usepackage{amsmath, mathtools, xfrac}	
\usepackage{amssymb, enumitem}	
\usepackage{longtable, multicol}
\usepackage{rotate, setspace}
\usepackage{lscape}
\usepackage{natbib, hyperref}
\usepackage{latexsym,verbatim,xcolor}
\newcommand{\apj}[1]{ApJ, }
\newcommand{\jcp}[1]{J. Chem. Phys., }
\newcommand{\mnras}[1]{MNRAS, }
\newcommand{\aj}[1]{AJ, }
\newcommand{\apjs}[1]{ApJS, }
\newcommand{\apjl}[1]{ApJ Letter, }
\newcommand{\aap}[1]{A\&A, }
\newcommand{\aaps}[1]{A\&A Suppl. Series, }
\newcommand{\araa}[1]{Annu. Rev. A\&A, }
\newcommand{\aaas}[1]{A\&AS, }
\newcommand{\apss}[1]{Ap\&SS }
\newcommand{\bain}[1]{Bul. of the Astron. Inst. of the Netherlands,}
\newcommand{\planss}[1]{Planetary and Space Science,}
\newcommand{\nat}[1]{Nature,}
\newcommand{\actaa}[1]{Acta Astronomica,}
\newcommand{\aapr}[1]{The Astronomy and Astrophysics Review,}
\newcommand{\memsai}[1]{Memorie della Societa Astronomica Italiana,}

\begin{document}\sloppy

\title{Enhancing SED-Based Astrometric, Photometric, and Kinematic Studies of SAI\,72 and SAI\,75 Using \textit{Gaia} DR3}

\author{A. Y. Alzhrani\textsuperscript{1}, A. A. Haroon\textsuperscript{1}, W. H. Elsanhoury\textsuperscript{2*}, and D. C. \c{C}{\i}nar\textsuperscript{3}}
\affilOne{\textsuperscript{1}Astronomy and Space Science Department, Faculty of Science, King Abdulaziz University, Jeddah, Saudi Arabia.\\}
\affilTwo{\textsuperscript{2}Physics Department, College of Science, Northern Border University, Arar, Saudi Arabia.\\}
\affilThree{\textsuperscript{3}Programme of Astronomy and Space Sciences, Institute of Graduate Studies in Science, Istanbul University, 34116, Beyaz{\i}t, Istanbul, Turkey.\\}

\twocolumn[{

\maketitle

\corres{elsanhoury@nbu.edu.sa}


\msinfo{19 FEB 2025}{21 APR 2025}

\begin{abstract}  
This study investigates the open clusters SAI\,72 and SAI\,75 using $Gaia$ DR3 data, employing the Automated Stellar Cluster Analysis (\text{ASteCA}) tool to determine their structural and fundamental properties, including center coordinates, size, age, distance, mass, luminosity, and kinematics. Based on membership probabilities ($P\geq50\%$), we identified 112 and 115 stars as probable members of SAI\,72 and SAI\,75, respectively. Radial density profile (RDP) analysis yielded cluster radii of 2.35 arcmin for SAI\,72 and 2.19 arcmin for SAI\,75. The spectral energy distribution (SED) fitting was performed to refine metallicity, distance, and color excess parameters, ensuring consistency within 1$\sigma$ of isochrone-based estimates. Isochrone fitting of the color-magnitude diagram (CMD) suggests ages of 316 Myr and 302 Myr, with corresponding distances of $3160 \pm 80$ pc and $3200 \pm 200$ pc. We derived their galactic positions, projected distances $(X_\odot,~Y_\odot)$, and vertical displacements $(Z_\odot)$. Mass function analysis estimates cluster masses of $612 \pm 174~M_\odot$ for SAI\,72 and $465 \pm 90~M_\odot$ for SAI\,75. Kinematic studies indicate that both clusters have reached dynamical equilibrium. The $AD$ diagram method provided convergent point coordinates of $(A,~D)_o = (97^{o}.016 \pm 0^{o}.09,~4^{o}.573 \pm 0^{o}.05)$ for SAI\,72 and $(99^{o}.677 \pm 0^{o}.10,~1^{o}.243 \pm 0^{o}.09)$ for SAI\,75. Orbital analysis confirms that both clusters follow nearly circular trajectories with low eccentricities and minor variations in apogalactic and perigalactic distances. Furthermore, we determine that SAI\,72 and SAI\,75 originated beyond the solar circle at \( R_{\rm Birth} = 10.825\pm0.068 \) and \( R_{\rm Birth} = 9.583\pm0.231 \) kpc, respectively. Their maximum heights above the Galactic plane, \(Z_{\rm max}\), are \(109 \pm 9\) pc for SAI\,72 and \(232 \pm 24\) pc for SAI\,75, reinforcing their classification as part of the young stellar disc population.  
\end{abstract}


}]


\doinum{12.3456/s78910-011-012-3}
\artcitid{\#\#\#\#}
\volnum{000}
\year{2025}
\pgrange{1--}
\setcounter{page}{1}
\lp{\#}

\section{INTRODUCTION}\label{sec1}

Open clusters (OCs) are fundamental building blocks of the Galactic disk, offering valuable insights into the formation and evolution of stellar populations. These gravitationally bound stellar systems originate from the same molecular cloud, ensuring that their member stars exhibit homogeneous properties in terms of age, chemical composition, distance, and kinematics, with mass being the primary differentiating factor among them \citep{LadaandLada2003}. Given their shared formation history, OCs serve as crucial astrophysical laboratories for investigating the underlying mechanisms that shape stellar evolution and Galactic structure.

The study of OCs extends far beyond individual cluster properties, as these systems act as tracers of Galactic evolution. Their spatial distribution and kinematics provide essential clues about the dynamical history of the Milky Way, particularly when analyzed in the context of their orbital motion within the Galactic potential \citep{Dias2005, Hao2021}. Recent research demonstrates that reconstructing the orbits of OCs enables the identification of their birthplaces and subsequent dynamical evolution, shedding light on the interaction between young stellar populations and large-scale Galactic structures such as spiral arms and resonances \citep{Monteiro2021}.

Spectral energy distribution (SED) analysis provides a crucial approach for deriving fundamental stellar parameters in OCs. By integrating multi-band photometric observations with high-precision spectroscopic constraints, SED fitting allows for a more accurate determination of essential parameters such as effective temperature, metallicity, interstellar extinction, and stellar radii \citep{Chunyan2024, Dursun2024, Oralhan2025}. The utilization of \textit{Gaia} astrometric data further enhances the reliability of membership probability assessments and color excess corrections, mitigating uncertainties introduced by field star contamination \citep{GaiaDR1, Castro2022}. Furthermore, recent advancements in Bayesian inference and machine learning techniques have significantly improved the ability to resolve parameter degeneracies, leading to more robust astrophysical parameter estimates \citep{Hunt2024}. 

The European Space Agency's (ESA) $Gaia$ mission, which commenced operations in December 2013 \citep{Prusti2016}, has significantly transformed our understanding of stellar populations, particularly in the study of OCs. The exceptional accuracy and depth of $Gaia$ astrometric and photometric data have been instrumental in the identification and detailed characterization of these stellar systems. Numerous new OCs have been uncovered through $Gaia$ observations \citep{Sim2019, Castro2022}, while its datasets have facilitated detailed investigations into their fundamental properties \citep{Yontan2022, Badawy2023}. Additionally, $Gaia$ contributions have been instrumental in refining membership probabilities and providing insights into the kinematic structure of these clusters \citep{Ferreira2021, Hao2022}.

\begin{figure}
\centering
\includegraphics[width=0.95\linewidth]{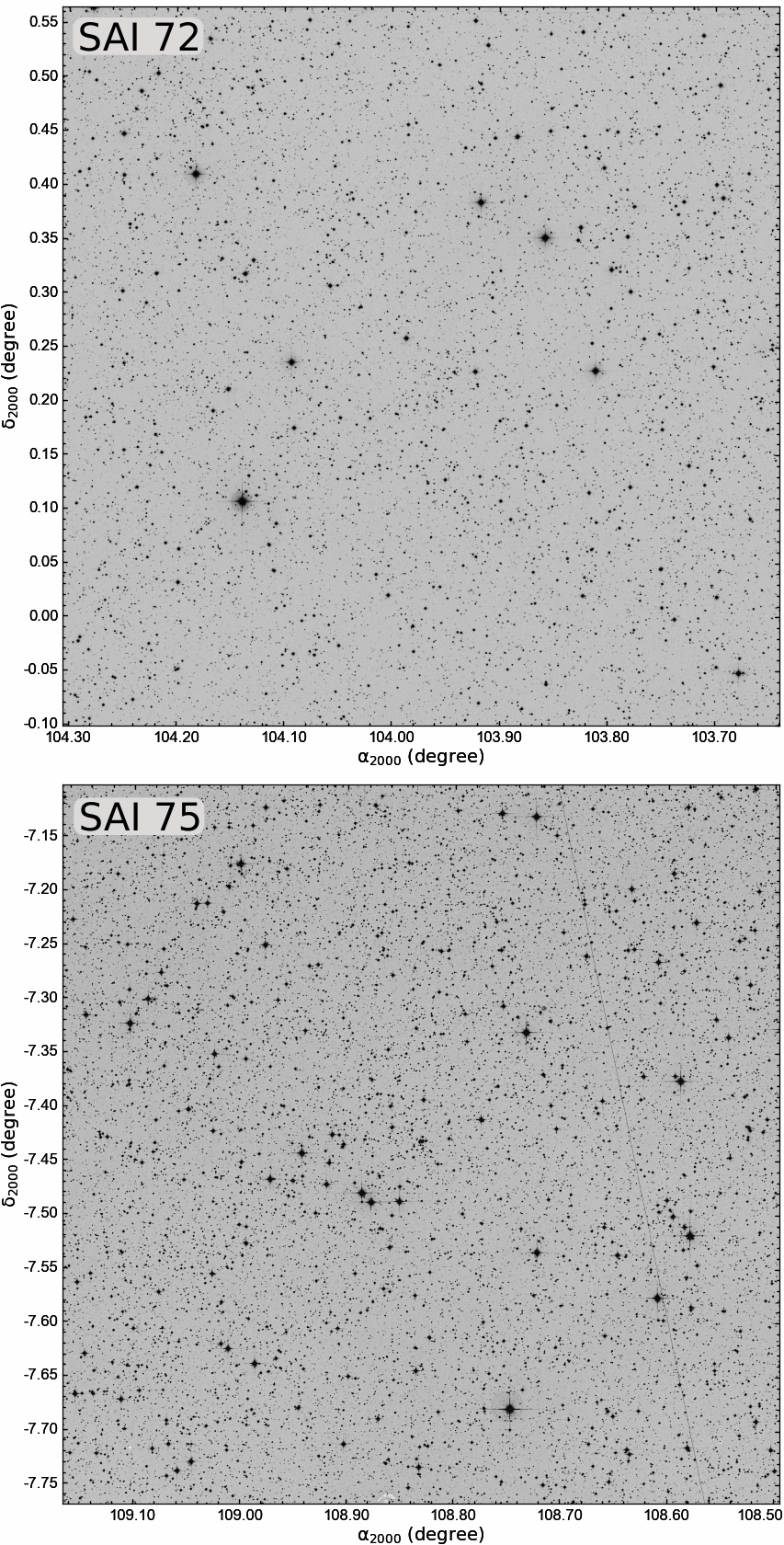}
\caption {Identification maps of SAI\,72 and SAI\,75.}
\label{Fig: 1}
\end{figure}

\begin{table*}
\setlength{\tabcolsep}{6pt}
\renewcommand{\arraystretch}{1}
\footnotesize
\centering
\caption{Astrophysical parameters for the SAI 71 and SAI\,75 OCs are provided in the table, which includes the color excess $E(B-V)$, distance ($d$), iron abundance [Fe/H], age, $\log(t)$, proper motion, and radial velocity. References are given in the 'Ref' column.}
  \begin{tabular}{cccccccl}
  \hline
    \multicolumn{8}{c}{SAI\,72}\\
    \hline
$E(B-V)$ & $d$ & [Fe/H] & $\log(t)$ &  $\langle\mu_{\alpha}\cos\delta\rangle$ &  $\langle\mu_{\delta}\rangle$ & $V_{\rm R}$ & Ref. \\
(mag) &  (pc)  & (dex) &  & (mas yr$^{-1}$) & (mas yr$^{-1}$) & (km s$^{-1})$ &      \\
\hline
0.82$\pm$0.06 & 3150$\pm$70 & --- & 8.50$\pm$0.20 & --- & --- & --- &   (01) \\
0.77 & 3239 & --- & 8.49 & --- & --- & --- &   (02) \\
0.34 & 5408 & --- & 8.85 & -0.512$\pm$0.17 & -0.267$\pm$0.422 & --- &   (03) \\
--- & 3696 & --- & --- & --- & --- & 46.79$\pm$3.26 &   (04) \\
--- & 4219$\pm$285& --- & 8.48 & -0.441$\pm$0.089 & -0.006$\pm$0.103 & 56.59$\pm$0.24 &   (05) \\
--- & 3747 & --- & --- & --- & --- & 46.79$\pm$3.26 &   (06) \\
--- & --- & --- & 8.48 & -0.441$\pm$0.138 & -0.006$\pm$0.102 & 54.60$\pm$1.00 & (07) \\
0.81$\pm$0.03 & 3159$\pm$80 & -0.05 & 8.50$\pm$0.01 & -0.43 & 0.05 & --- &   (12) \\
\hline
    \multicolumn{8}{c}{SAI\,75}\\
\hline
0.23$\pm$0.03 & 2800$\pm$330 & --- & 8.60 & --- & --- & --- &   (01) \\
--- & --- & 0.10 & --- & --- & --- & --- &   (08) \\
 0.23 & 3522$\pm$420 & --- & 7.81 & -1.31$\pm$0.09 & 1.36$\pm$0.06 & --- &  (09) \\  
 --- & 2870 & --- & --- & -1.29 & 1.35 & 106.90 &  (10) \\
 0.25$\pm$0.02 & 3226$\pm$385 & 0.07 & 8.46 & --- & --- & --- &  (11) \\
0.29$\pm$0.03 & 3200$\pm$200 & 0.10 & 8.48$\pm$0.07 & -0.26 & 0.22 & --- &   (12) \\
  \hline
    \end{tabular}%
    \\
\raggedright
{\scriptsize (01)~\citet{Glushkova2010} (02)~\citet{Kharchenko2016}, (03)~\citet{Loktin2017}, (04)~\citet{Soubiran2018}, (05)~\citet{Hao2021}, (06)~\citet{Tarricq2021}, (07)~\citet{Liu2023}, (08)~\citet{Yadav2014},(09)~\citet{Castro2022}, (10)~\citet{Hunt2024}, (11)~\citet{Cavallo2024} (12)~This study. }
\label{tab:literature}%
\end{table*}%

The clusters examined in this study were selected from a catalogue developed by researchers at the Sternberg Astronomical Institute (SAI), Russia \citep{koposov2008,Glushkova2010}. This catalogue, which is based on observations from the Two Micron All Sky Survey (2MASS) in the $JHK_s$ infrared bands \citep{Skrutskie2006}, provides the initial astrophysical parameters of identified OCs. In this study, we selected two open clusters, SAI 72 and SAI 75, that belong to the same Galactic region. These clusters, despite being catalogued in large-scale surveys, have not been the focus of detailed individual studies. While previous research has primarily included them in broad analyses of open cluster populations, in-depth investigations into their structural properties, kinematics, and evolutionary states remain limited. Among the selected clusters, SAI\,75 has been previously examined by \cite{Yadav2014}, who employed $UBVI$ photometric observations conducted with the 104-cm Sampurnanand Telescope at ARIES, Nainital, India. Their analysis provided key insights into the cluster’s characteristics, including its age, metallicity, distance, and \text{color} excess, and their findings confirmed that its stars follow the standard interstellar extinction law.

The fundamental parameters of the SAI\,72 and SAI\,75 were obtained from the OCs catalogue\footnote{\href{http://ocl.sai.msu.ru/catalog/}{http://ocl.sai.msu.ru/catalog/}}. The equatorial coordinates (J2000) of SAI\,72 are $\alpha = 6^{\text{h}} 55^{\text{m}} 48^{\text{s}}$, $\delta = +00^{\circ} 13' 37''$, while SAI\,75 is located at $\alpha = 7^{\text{h}} 15^{\text{m}} 23^{\text{s}}$, $\delta = -07^{\circ} 25' 20''$. Their respective Galactic coordinates are $(l,~ b) = (213.228^{\circ}, +1.0750^{\circ})$ for SAI\,72 and $(222.266^{\circ}, +1.80^{\circ})$ for SAI\,75. Their Digitized Sky Survey (DSS) images are shown in Figure \ref{Fig: 1}. The \text{color excesses} were determined as $E(B-V) = 0.82 \pm 0.06$ for SAI\,72 and $E(B-V) = 0.23 \pm 0.03$ for SAI\,75, indicating differences in their line-of-sight extinction. Additionally, the logarithm of their ages was found to be $\log(t) = 8.5 \pm 0.2$ for SAI\,72, while SAI\,75 has an upper age limit of $\log(t) < 8.6$, suggesting that both clusters are relatively young. The results of SAI\,72 and SAI\,75 as reported in the literature are presented in Table \ref{tab:literature}.

This paper is organized as follows: The details of the data used in this study are presented in Section \ref{sec2} We discuss the results of the $Gaia$ DR3 data analysis of the selected open clusters, including the estimation of their astrometric, photometric, dynamical, and kinematic parameters in Section \ref{sec3} Finally, in section \ref{sec4} we present our conclusion about the main result.

\section{DATA AND \textit{Gaia} DR3}
\label{sec2}

\subsection{Photometric and Astrometric Data}

For this study, we utilized astrometric and photometric data from the $Gaia$ DR3 catalog \citep{GaiaDR3} to analyze the properties of the target clusters. The $Gaia$ DR3 dataset provides high-precision astrometric measurements, including equatorial coordinates ($\alpha,~\delta$), trigonometric parallax ($\varpi$), and proper motion components ($\mu_{\alpha}\cos\delta,~\mu_{\delta}$), along with photometric observations in three bands; $G$ (330–1050 nm), $G_{\rm BP}$ (330–680 nm), and $G_{\rm RP}$ (630–1050 nm).

For our sample, we extracted data from regions of 40 arcminute centered on SAI\,72 and SAI\,75. The compiled dataset includes three-band photometry ($G$,~$G_{\rm BP}$,~$G_{\rm RP}$), along with key astrometric parameters such as coordinates, proper motions, and parallaxes, along with their associated uncertainties. A significant step in the realm of OC studies is the identification of true members and the distinction between these and field star contamination. This is due to the fact that foreground and background stars have the capacity to introduce significant biases when determining cluster properties \citep{Carraro2008}.

\subsection{Photometric Completeness Limit and Errors}

\begin{figure*}
\centering
\includegraphics[width=0.9\linewidth]{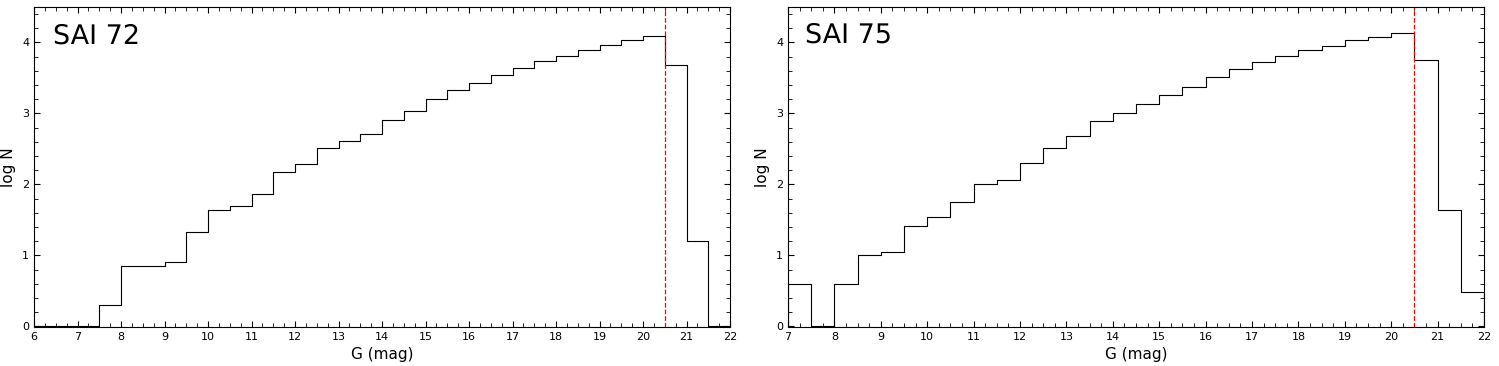}
\caption{The distribution of stars along the direction of SAI\,72 and SAI\,75 is shown across various $G$-magnitude ranges, with the photometric completeness limit indicated by a red dashed line.}
\label{Fig: Comp}
\end{figure*}

Accurate determination of the structural and astrophysical parameters of SAI\,72 and SAI\,75 necessitates establishing the photometric completeness limit by analyzing the distribution of stars as a function of $G$ magnitude. The completeness limit is identified as the magnitude at which the number of detected stars increases up to a peak before declining due to observational limitations. Based on the histogram in Figure~\ref{Fig: Comp}, the photometric completeness limits for SAI\,72 and SAI\,75 are determined as $G=20.5$ mag. To ensure the reliability of subsequent analyses, stars fainter than these limits were excluded. 

Photometric uncertainties provided in {\it Gaia} DR3 were interpreted as internal errors, representing the instrumental uncertainties associated with stellar magnitudes. Consequently, these uncertainties were incorporated into the analysis as intrinsic measurement errors. The mean errors for $G$ magnitudes and $G_{\rm BP}-G_{\rm RP}$ color indices were calculated within distinct $G$ magnitude intervals, with the results summarized in Table \ref{tab:photometric_errors}.

\begin{table}
\centering
\footnotesize
\caption{The mean internal photometric uncertainties for SAI\,72 and SAI\,75 in the $G$ and $G_{\rm BP} - G_{\rm RP}$ bands are calculated across different $G$-magnitude bins.}
\begin{tabular}{cccc}
\hline
$G$ & $N$ & $\sigma_G$& $\sigma_{G_{\rm BP}-G_{\rm RP}}$ \\
(mag) & & (mag) & (mag) \\
\hline
\multicolumn{4}{c}{SAI\,72} \\
\hline
 (06, 14] &  1554 & 0.003 & 0.005 \\
 (14, 15] &  1563 & 0.003 & 0.005 \\
 (15, 16] &  3151 & 0.003 & 0.006 \\
 (16, 17] &  5499 & 0.003 & 0.010 \\
 (17, 18] &  8831 & 0.003 & 0.020 \\
 (18, 19] & 13035 & 0.004 & 0.047 \\
 (19, 20] & 18699 & 0.005 & 0.108 \\
 (20, 21] & 22135 & 0.011 & 0.221 \\
 (21, 23] &   468 & 0.028 & 0.388 \\
\hline
\multicolumn{4}{c}{SAI\,75} \\
\hline
(06, 14] & 1730 & 0.003 & 0.001 \\
(14, 15] & 2073 & 0.003 & 0.001 \\
(15, 16] & 3624 & 0.003 & 0.000 \\
(16, 17] & 6531 & 0.003 & 0.002 \\
(17, 18] & 10744 & 0.003 & 0.012 \\
(18, 19] & 15453 & 0.004 & 0.019 \\
(19, 20] & 21296 & 0.006 & 0.041 \\
(20, 21] & 24662 & 0.013 & 0.516 \\
(21, 23] & 600 & 0.030 & 4.686 \\
\hline
\end{tabular}
\label{tab:photometric_errors}
\end{table}

\subsection{Membership Analysis}

To investigate the physical properties of the clusters, we employed the Automated Stellar Cluster Analysis (\text{ASteCA}) software \citep{Perren2015}, which facilitates the determination of fundamental parameters such as age, metallicity, distance, and interstellar extinction. This tool applies statistical techniques to model cluster distributions and estimate their inherent uncertainties. A detailed explanation of \text{ASteCA}'s methodology can be found in \cite{Perren2015} and its official online documentation\footnote{\url{http://asteca.github.io/}}.

To address this, \text{ASteCA} employs Bayesian inference to assign membership probabilities to stars. This method systematically compares the kinematic and spatial properties of stars within the cluster region to those in nearby field areas, allowing for a robust distinction between true members and contaminants. Stars with membership probabilities of at least $P \geq 50\%$ are considered likely cluster members \citep{Perren2020}. Applying this technique, we identified 112 probable members for SAI\,72 and 115 for SAI\,75, providing a reliable foundation for further analysis.

\section{RESULTS AND DISCUSSION}\label{sec3}

\subsection{Center and Radius Assignment }
\begin{figure}
\centering
\includegraphics[width=0.7\linewidth]{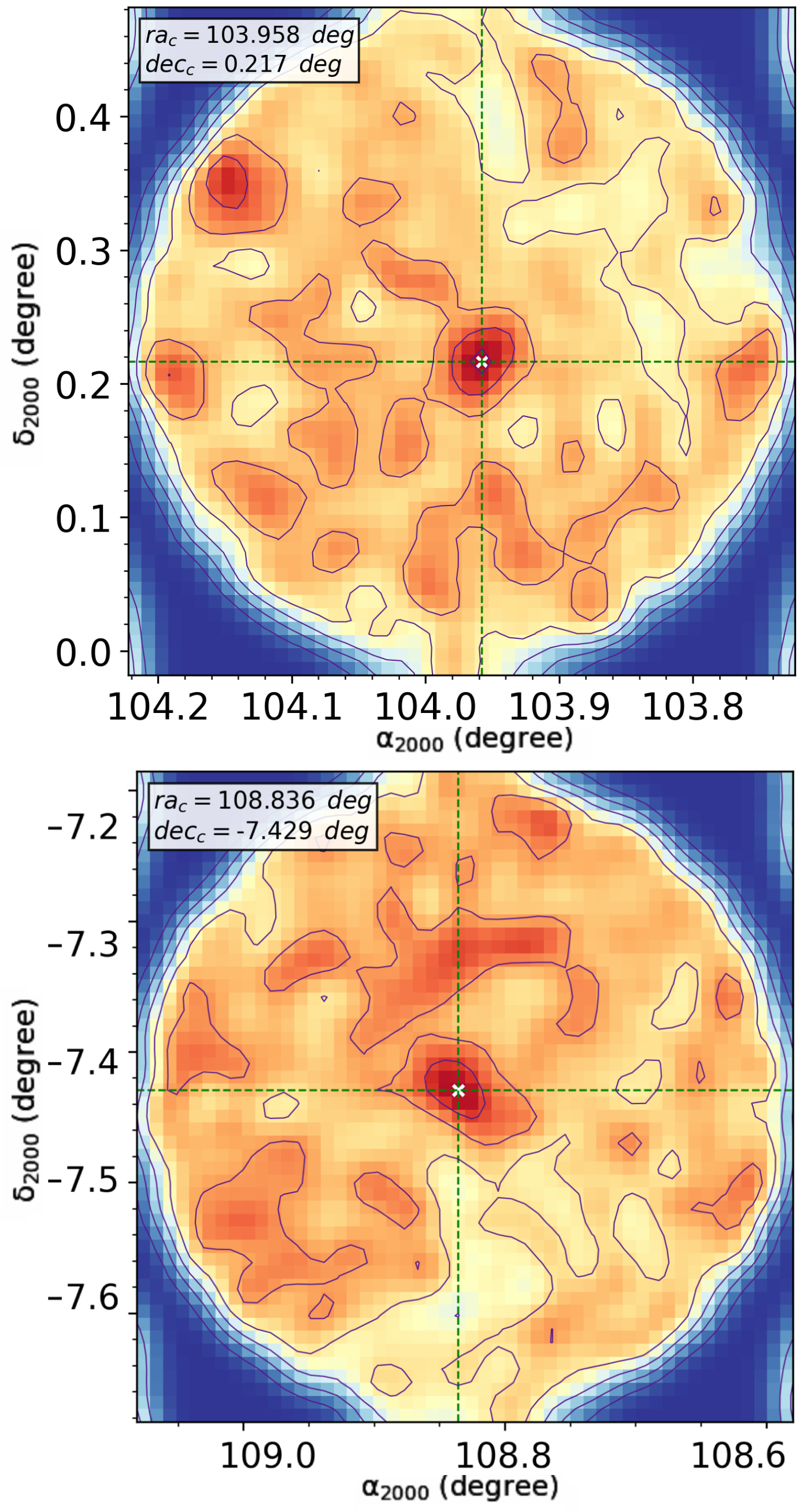}
\caption{The structural density contour maps for SAI\,72 are shown in the upper panel, and for SAI\,75 in the lower panel.}
\label{Fig: 2}
\end{figure}

\begin{figure*}
\includegraphics[width=0.99\linewidth]{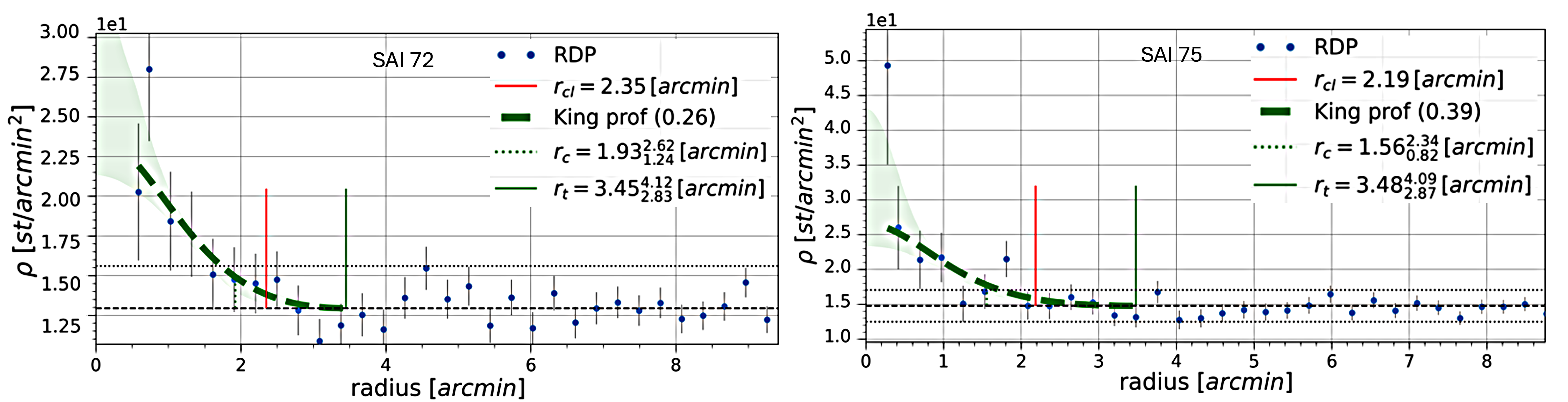}
\caption{The King profile is fitted to the observed surface density to determine the RDP parameters for the clusters. The blue dots represent the RDPs calculated using the \text{ASteCA} code. The green dashed curve and shaded region depict the King density profile. The black dotted and dashed curves indicate $\rho_{\circ}$ and $\rho_{\text{bg}}$, respectively. The vertical thin lines mark the core, limiting, and tidal radii, as labeled in the figure key.}
\label{Fig: 3}
\end{figure*}

Accurately identifying the central coordinates of star clusters is essential for determining key physical parameters, including age, distance, and interstellar extinction. In this study, we employed the kernel density estimation (KDE) method within the \text{ASteCA} framework to locate the cluster centers. The peak of the stellar density distribution was adopted as the central position of each cluster. Figure \ref{Fig: 2} illustrates the density contours for both clusters. As a result of this analysis, the central coordinates in right ascension and declination were determined to be ($103^{\circ}.958, 0^{\circ}.217$) for SAI\,72 and ($108^{\circ}.836, -7^{\circ}.429$) for SAI\,75. The final coordinates, given in both equatorial $(\alpha,~\delta)$ and galactic $(l,~b)$ reference frames, are summarized in Table \ref{Tab: 2}. Our findings indicate a strong agreement with the cluster positions previously reported by \cite{Glushkova2010}. The cluster radii were estimated by concentrating the radial density profile (RDP) as shown in Figure \ref{Fig: 3}. By \text{ASteCA} code, we obtained the observed RDPs for the two clusters and fitted them with a King’s profile \citep{King1962} as in Equation (\ref{Eq.1}).
\begin{eqnarray} \rho(r)= \rho_{\rm bg}+\dfrac{\rho_o} {1+(r/r_{\rm c})^2}, \label{Eq.1} \end{eqnarray}
where $\rho_o$ and $\rho_{bg}$ represent the central and background stellar densities, respectively. RDP was derived by constructing a grid-based two-dimensional histogram and analyzing concentric square annuli around the center of cluster. The \text{ASteCA} algorithm defines the limiting radius $(r_{cl})$ as the point at which the density profile merges with the background field level. The estimated limiting radii for SAI\,72 and SAI\,75 are 2.35 arcminutes and 2.19 arcminutes, respectively. Additionally, the core radius $(r_c)$ is defined as the radial distance where the stellar density $\rho(r)$ falls to half of the central density $\rho_o$.

\begin{table*}
\centering
\caption{The calculated coordinates of the new center positions of the clusters are presented in both equatorial ($\alpha$, $\delta$) and Galactic ($l$, $b$) systems.}
\label{Tab: 2}
\begin{tabular}{lcccc} 
\hline
                 & \multicolumn{2}{c}{\textbf{Equatorial coordinates}}                   & \multicolumn{2}{c}{\textbf{Galactic coordinates}}  \\ 
\hline
\textbf{Cluster} & \textbf{$\alpha$}                         & \textbf{$\delta$}         & \textbf{$l$}     & \textbf{$b$}                    \\ 
\hline\hline
SAI\,72            & $6^\text{h}\;55^\text{m}\;50.39^\text{s}$ & $+00^\circ\;12'\;01.58''$ & $213^\circ.2554$ & $+1^\circ.0709$                 \\
SAI\,75            & $7^\text{h}\;15^\text{m}\;18.27^\text{s}$ & $-07^\circ\;26'\;16.19''$ & $222^\circ.2722$ & $+1^\circ.8743$                 \\
\hline
\end{tabular}
\end{table*}

To further characterize the structural properties of these clusters, we computed the density contrast parameter ($\delta_{\rm c}$) and the concentration parameter ($C$). The density contrast, given by $\delta_{\rm c} = 1 + \rho_o / \rho_{\rm bg}$, quantifies the cluster's relative density in comparison to the surrounding stellar field. Higher values of $\delta_{\rm c}$, within the range $7 \lesssim \delta_{\rm c} \lesssim 23$, indicate a higher degree of central concentration, which is typical for densely packed clusters. The concentration parameter, defined as $C = (r_{\rm cl}/r_{\rm c})$, measures the compactness of the core region \citep{king1966}. In our analysis, we obtained values of $C = 1.22$ for SAI\,72 and $C = 1.40$ for SAI\,75. The estimated radii for SAI\,75 in our study are consistent with the values reported by \citep{Yadav2014}. Additionally, Table \ref{Tab: 3} provides a comparative overview of our RDP parameters alongside those obtained by \citep{Yadav2014}.

\begin{center}
\begin{table}[t]%
\centering
\footnotesize
\caption{The structural parameters of the clusters under study, derived from the King profile fit, are presented. These are compared with the parameters for SAI\,75 from present work (denoted as p.w.) and those reported by \cite{Yadav2014} (denoted as *).}
\begin{tabular}{lccc}
\hline
\textbf{Parameters} & \textbf{SAI\,72}  & \textbf{SAI\,75} & \textbf{Reference} \\
\hline
\hline
$r_{\rm c}$ (arcmin) & $1.93^{+2.62}_{-1.24}$ & $1.56^{+2.34}_{-0.82}$ & p.w. \\
               &                        & $1.20 \pm 0.50$          & (*) \\
$r_{\rm cl}$ (arcmin) & 2.35  & 2.19 & p.w.  \\
 & & 2.5& (*)\\
 $r_{\rm t}$ (arcmin) & $3.45^{4.12}_{2.83}$  & $3.48^{4.09}_{2.87}$ & p.w. \\
$\rho_o$ (stars arcmin$^{-2}$) & 15.72  & 17.50  & p.w.  \\
& &  $6.10 \pm1.40$& (*)\\
$\rho_{\rm bg}$ (stars arcmin$^{-2}$) & 12.93  & 14.76  & p.w.  \\
& & $4.20\pm0.50$&(*)\\
$\delta_{\rm c}$ & 2.22 & 2.19 & p.w. \\
& & $\sim 2.50$& (*)\\
$C$  & 1.22  & 1.40  & p.w.  \\
\hline
\end{tabular}
\label{Tab: 3}
\end{table}
\end{center}

\subsubsection{Astrometric Parameters \\}
\begin{figure}
\centering
\includegraphics[width=0.95\linewidth]{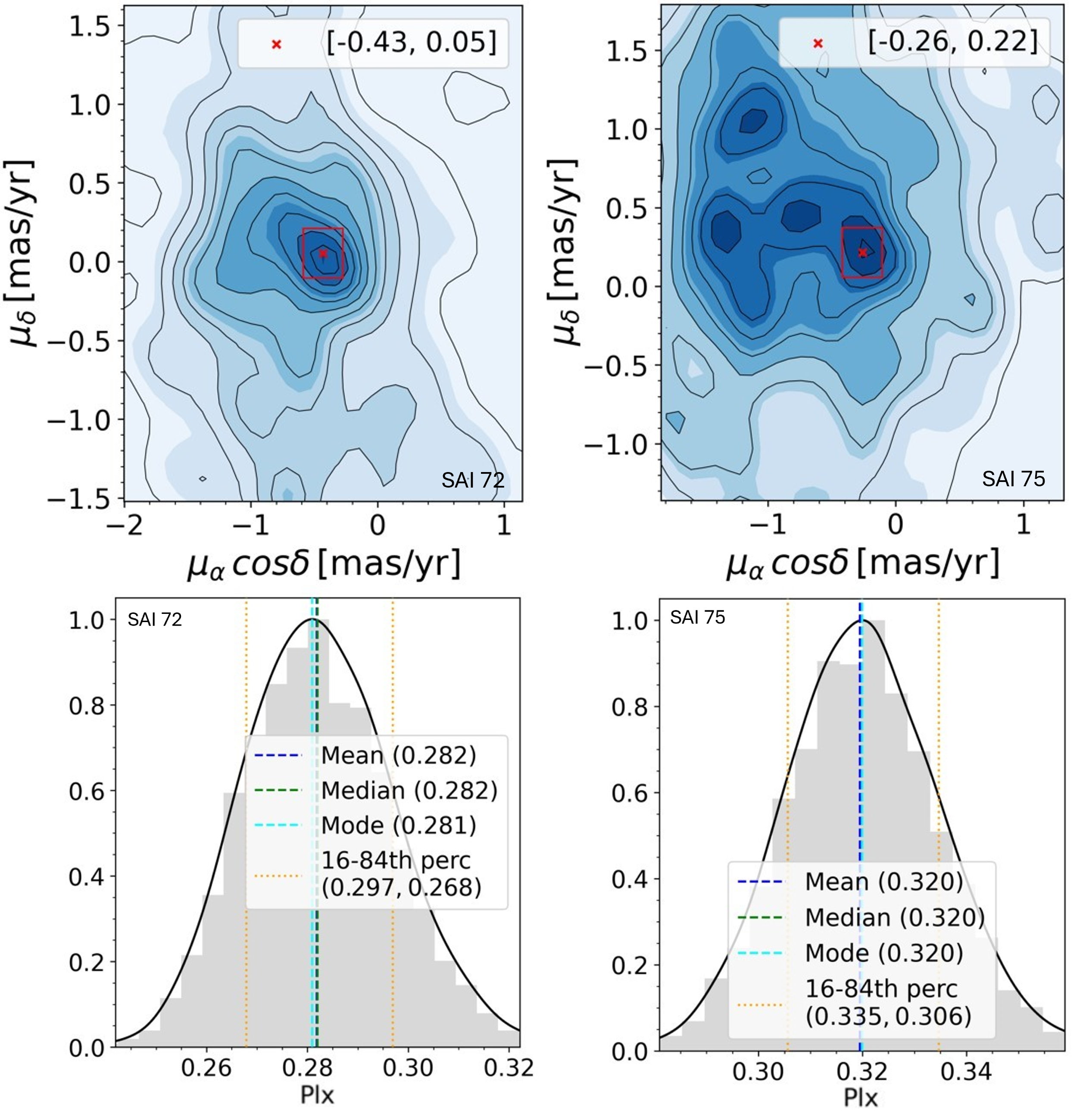}
\caption{For the two clusters, the upper panel presents the spread of the mean proper motion in both right ascension and declination directions, while the lower panel presents the parallax gaussian distribution.}
\label{Fig: 4}
\end{figure}

The stellar distribution of each cluster was plotted in the proper motion space $(\mu_{\alpha}~cos\delta,~\mu_{\delta})$, as depicted in the upper panels of Figure \ref{Fig: 4}. The mean proper motion components of the cluster were subsequently determined by fitting a Gaussian distribution along the corresponding directions. In the lower panel of Figure \ref{Fig: 4} histograms are presented, which display the trigonometric parallax distribution of the candidate members of the clusters. The black line on the histograms is the Gaussian fitting from which the mean trigonometric parallax of each cluster was estimated. Using the trigonometric parallax estimated values, we determined the distances to each cluster. The calculated distances ($d_{\varpi}$) are as follows: 3548 $\pm$ 60 and 3130 $\pm$ 56 pc for SAI\,72 and SAI\,75, respectively. The mean trigonometric parallax values and mean proper motion components for the clusters were presented in Table \ref{Tab: 4}. 

\subsection{Photometric Analysis}
\subsubsection{Age and Distance \\}
\begin{figure}
\centering
\includegraphics[width=0.9\linewidth]{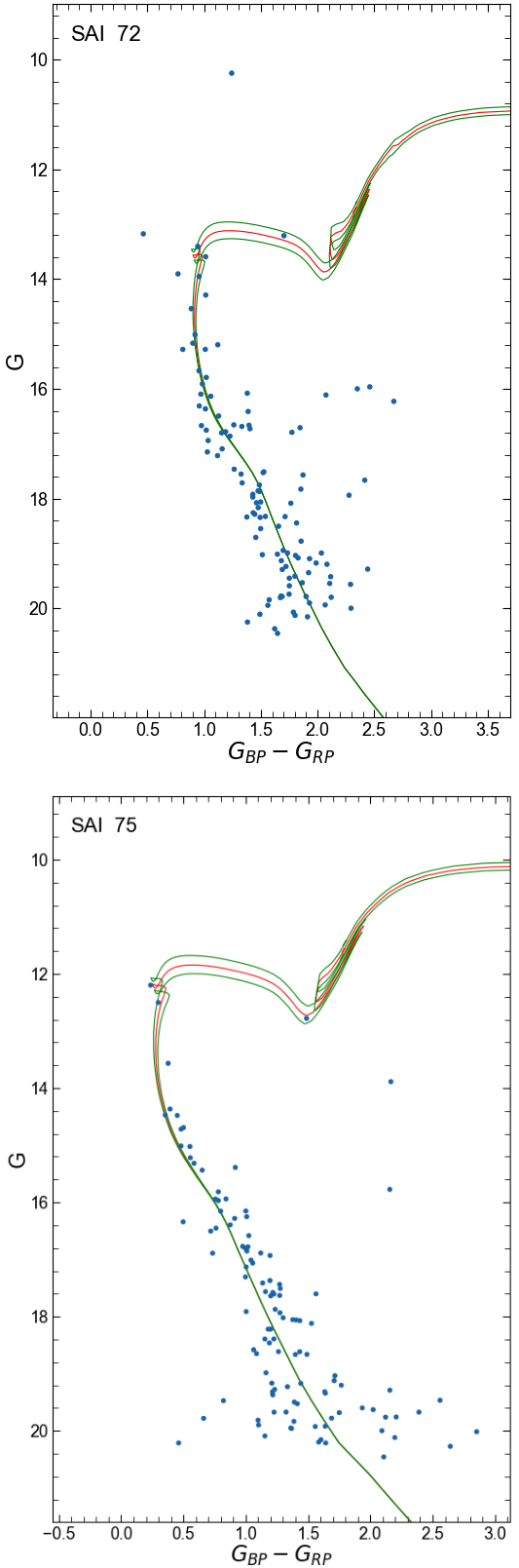}
\caption{The CMDs for the member stars of each cluster are shown, with the red lines representing the best-fit isochrones, with the associated error margins representing in green lines for SAI\,72 and SAI\,75.}
\label{Fig: 5}
\end{figure}

The color-magnitude diagram (CMD) serves as a fundamental diagnostic tool for deriving essential cluster parameters, including age, distance modulus, \text{color} excess, and metallicity. These parameters can be estimated using the \text{ASteCA} package, which employs an isochrone-fitting technique based on synthetic CMD generation and optimization via a genetic algorithm. To determine these properties, we analyzed the photometric magnitudes ($G$,~$G_{\rm BP}$,~$G_{\rm RP}$) of member stars in the selected clusters. \text{ASteCA} was applied to fit theoretical isochrones from the PARSEC v1.2S model suite \citep{Bressan2012}, as demonstrated in Figure \ref{Fig: 5}. The \text{color} excess, $E(G_{\rm BP}-G_{\rm RP})$, was determined using the empirical relation $E(G_{\rm BP}-G_{\rm RP}) = 1.289 \times E(B-V)$ \citep{Elsanhoury2022}. Additionally, extinction correction was applied using the relation $A_{\rm G} = 2.74 \times E(B-V)$ \citep{Casagrande2018, Zhong2019}. The CMD-based distance estimates were cross-verified with astrometric distances derived from cluster parallax values ($d_{\varpi}$).

For SAI\,72, we determined a metallicity of $0.01349 \pm 0.0023$, while the logarithmic age remains consistent with previous studies \citep{Glushkova2010}, at approximately 8.5. Furthermore, both the derived distance modulus and distance values align closely with earlier findings. Regarding SAI\,75, our analysis yields a solar metallicity of $0.01920 \pm 0.00028$, with a slightly higher logarithmic age ($8.48 \pm 0.066$) compared to the value reported by \cite{Yadav2014}, $7.95 \pm 0.05$. The distance modulus remains in agreement with prior works \citep{Glushkova2010, Yadav2014}. However, the \text{color} excess shows minor discrepancies, and our study estimating $0.292 \pm 0.034$ mag, as opposed to $0.23 \pm 0.03$ mag from \cite{Glushkova2010} and $0.34 \pm 0.05$ mag from \cite{Yadav2014}. The \text{color} excess is found to be $0.376 \pm 0.043$. The distance to SAI\,75 is estimated to be around ($3200 \pm 200$) pc.

Table \ref{Tab: 4} summarizes these fundamental cluster parameters, emphasizing how the present analysis refines previous estimates by incorporating the latest observational data and modeling techniques. While our results generally \text{are consistent with previous studies}, minor variations arise due to methodological differences and dataset selection. These deviations highlight the ongoing need for precise parameter determination in stellar population studies. The Galactocentric distance of each cluster ($R_{\rm gc}$) was computed using the  $\sqrt{R_{\circ}^{2} + (d \cos b)^2 - 2 R_{\circ} d \cos b \cos l}$ relation, where $R_{\circ}$, the Sun's distance from the Galactic center, is taken as $8.20 \pm 0.10$ kpc \citep{Bland2019}. Here, $d$ represents the cluster's distance, and $b$ and $l$ denote its Galactic latitude and longitude, respectively. The Cartesian coordinates $(X_\odot,~Y_\odot,~Z_\odot)$ in the Galactic frame were computed using $X_\odot = d \cos b \cos l, \quad Y_\odot = d \cos b \sin l, \quad Z_\odot = d \sin b$ \citep{Elsanhoury2022}, facilitating a deeper understanding of their spatial distribution within the Milky Way.

\begin{table*}[htbp]
\centering
\caption{A comparison of the astrophysical and photometric parameters for SAI\,72 and SAI\,75 obtained in this study with those reported in the literature.}
\renewcommand{\arraystretch}{1.3} 
\begin{tabular}{lcccl}
\hline
\textbf{Parameters}   & \textbf{SAI\,72} & \textbf{SAI\,75} &  \textbf{Ref.} \\ 
\hline
\hline
No. of members& 112 & 115 & This study\\
$\mu_{\alpha}cos\delta$ (mas yr$^{-1}$)&  $-0.43$ $\pm$ 0.05
 & $-0.26$ $\pm$ 0.05 & This study\\
$\mu_{\delta}$ (mas yr$^{-1}$)&  0.05 $\pm$ 0.01 & 0.22 $\pm$ 0.04 & This study\\
$\varpi$ (mas) &  0.282 $\pm$ 0.05 & 0.320 $\pm$ 0.06 &This study \\
$\text{d}_{\varpi} $ (pc) &3548 $\pm$ 60 & 3130 $\pm$ 56 &This study\\
$Z$ &  $0.01349\pm0.0023$ & $0.01920\pm0.00028$ &This study \\
&--- &0.019& \cite{Yadav2014}\\
{\text{log}(\text{age yr$^{-1}$})}    & $ 8.50\pm 0.013 $ & $8.48\pm 0.066$& This study    \\ 
                      & $8.50\pm 0.20$  &  $<8.60$ & \cite{Glushkova2010} \\ 
                      &      ---   &  $7.95\pm 0.05$& \citet{Yadav2014}\\                     
{$E (B-V)$ (mag)} 
                      & 0.81$\pm$0.029 & $0.29\pm 0.034 $ & This study   \\ 
                      & $0.82\pm 0.06$ & $0.23 \pm 0.03$ & \cite{Glushkova2010} \\ 
                      & --- & $0.34 \pm 0.05$ & \cite{Yadav2014} \\

{$E(\text{G}_\text{BP}- \text{G}_\text{RP})$} 
                     & $1.039\pm0.037$ & $0.376\pm0.043$ &This study\\
{$A_{\text{G}}$} & 2.208 & 0.80 &This study \\

{$(m-M)_{o}$} (mag) & $12.498\pm 0.055$ & $12.54\pm 0.12$ &This study \\
                     & $12.49 \pm 0.05$ & $12.24\pm 0.24$ 
 &\cite{Glushkova2010} \\
 &--- & $12.7 \pm 0.2$& \cite{Yadav2014} \\
                    
{$d$ (pc)} & $ 3160 \pm 80$ & $ 3200\pm200$ & 
                        This study \\
                        & $3150 \pm 70$ & $2800\pm330$& \cite{Glushkova2010} \\
                        & ---& $3500\pm300$& \cite{Yadav2014} \\

{$R_\text{gc}$ (kpc)} & 10.9 $\pm$ 0.11 & 10.783 $\pm$ 0.10 &This study \\        
                        &11.609 & --- &\cite{Cantat-Gaudin2018} \\
                        & ---&11.30& \cite{Yadav2014} \\

{$X_\odot$ (kpc)} &$-2.641$ $\pm$ 0.05 & $-2.367$ $\pm$ 0.05 &This study \\
                                &$-3.091$& --- &\cite{Cantat-Gaudin2018}\\
                                & ---&$-2.350$& \cite{Yadav2014} \\
{$Y_\odot$ (kpc) }& $-1.732$ $\pm$ 0.04 & $-2.151$ $\pm$ 0.05&This study \\
                                 & $-2.025$ & &\cite{Cantat-Gaudin2018}\\
                                 & ---&$2.600$& \cite{Yadav2014} \\
{$Z_\odot$ (kpc)} &0.060 $\pm$ 0.01 & 0.105 $\pm$ 0.01 &This study \\
                                 &0.070&--- &\cite{Cantat-Gaudin2018}\\
                                   &--- &0.115& \cite{Yadav2014} \\
                                 \hline

\end{tabular}
\label{Tab: 4}
\end{table*}

\subsubsection{Spectral Energy Distribution\\}
We performed a comprehensive spectral energy distribution (SED) fitting for SAI\,72 and SAI\,75 OCs using {\sc ARIADNE}, a specialized PYTHON package designed for Bayesian analysis of stellar photometry. This tool enables the fitting of SEDs by employing a range of precomputed stellar atmosphere models, which have been convolved with various publicly available broad-band filter sets \citep{Vines2022}. To efficiently explore the multi-dimensional parameter space, {\sc ARIADNE} utilizes the nested sampling (NS) algorithm, implemented through dynesty, which simultaneously estimates the Bayesian evidence for different models while generating the corresponding posterior distributions \citep{Higson2019, Speagle2020}. The fundamental stellar parameters, including effective temperature, surface gravity ($\log g$), metallicity [Fe/H], visual extinction ($A_{\rm V}$), and stellar radius, were inferred by incorporating $Gaia$ EDR3 distance measurements from \citet{BailerJones2021}. Additionally, extinction values along the line of sight were determined from the SED dust map \citep{Schlafly2011}, ensuring a consistent approach to \text{color} excess corrections. The final parameter set for each star was derived through Bayesian model averaging, a technique that weights individual models based on their respective statistical evidence, thus mitigating biases from any single model assumption.

\begin{figure*}
\centering
\includegraphics[width=0.85\linewidth]{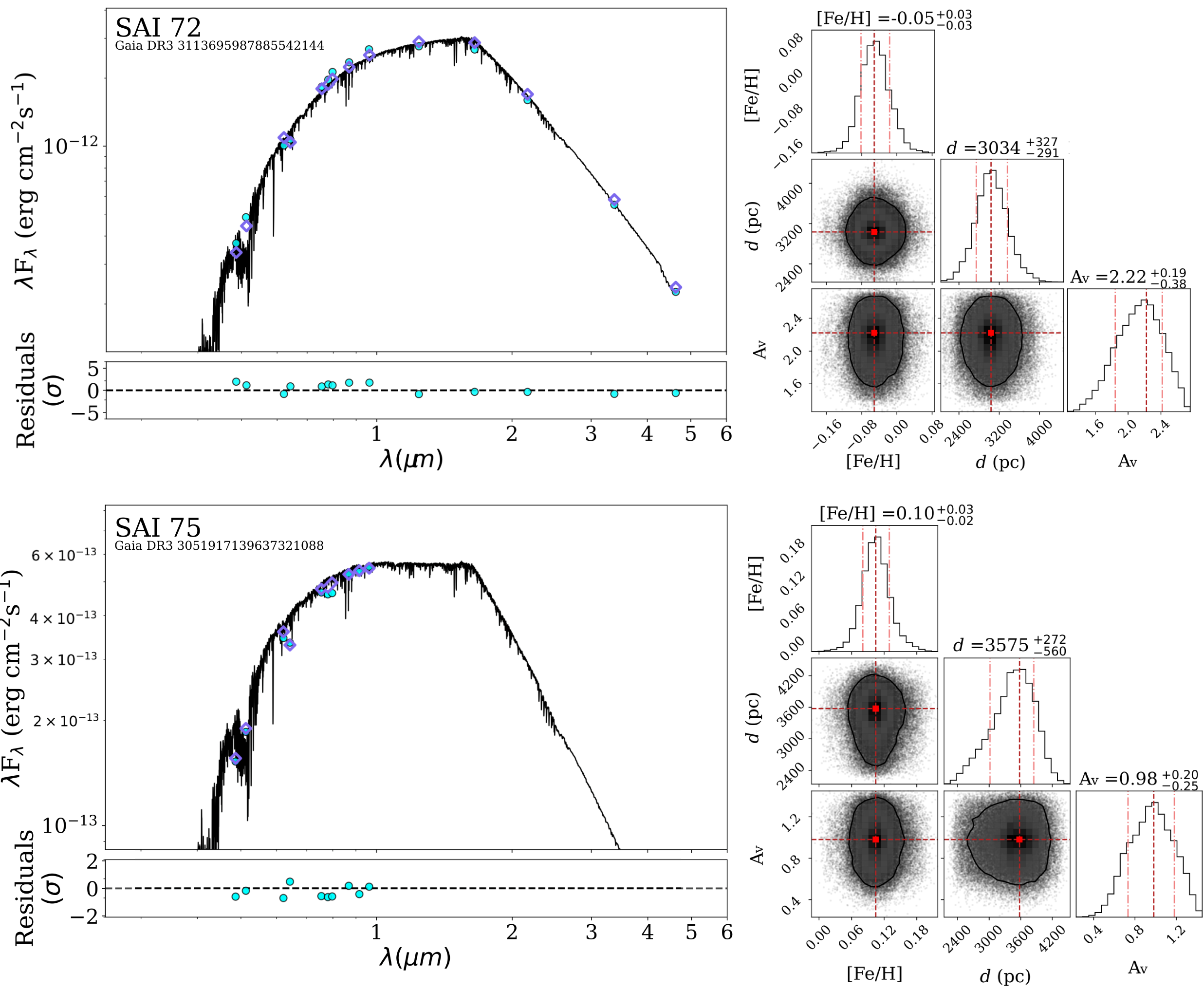}
\caption{The SED plots for two star members of the SAI\,72 and SAI\,75 OCs are presented in the left panels, along with the histograms and distribution plots of the optimal astrophysical parameter solutions shown in the right panels.}
\label{Fig: SED_corner}
\end{figure*}

Astrometric reliability assessments in the $Gaia$ Data \citep{GaiaDR2, GaiaDR3} include the Renormalized Unit Weight Error (RUWE), a key metric used to evaluate the quality of the astrometric solution for individual sources. For a well-behaved single star, the RUWE value is expected to be close to 1.0, reflecting a good fit between the model and the observed astrometric data. However, when RUWE exceeds 1.4, it often indicates deviations from the single-star assumption, potentially due to binarity, instrumental effects, or other complexities affecting the astrometric measurements.

We only excluded stars based on the RUWE threshold (RUWE > 1.4) during the SED analysis. Since the ARIADNE code we used for the SED analysis works best for single stars, we did not include stars that could potentially be binaries at this stage. After evaluating the RUWE values of the cluster members, 6 stars from SAI,72 and 2 stars from SAI,75 were identified as having RUWE values above the threshold and were subsequently removed from the SED analysis. After this selection process, we retained 106 members from SAI,72 and 113 members from SAI,75, for which we conducted detailed SED analyses. For the stellar populations of SAI\,72 and SAI\,75, we utilized the Phoenix V2 stellar atmosphere grid \citep{Husser2013}, which encompasses the broadest range of parameters. 

The results of the SED analyses of one dwarf star each for SAI 72 and SAI 75 are shown in Figure \ref{Fig: SED_corner}, with the corner lines indicating the agreement of the main astrophysical parameters. This is confirmed by the residual distributions in the bottom panel of the SED plots for each star. Furthermore, the corner lines in the right panel of the SED distributions show that there is \text{no strong correlation observed between parameters}, with uncertainties remaining within acceptable limits.

\begin{figure*}
\centering
\includegraphics[width=0.85\linewidth]{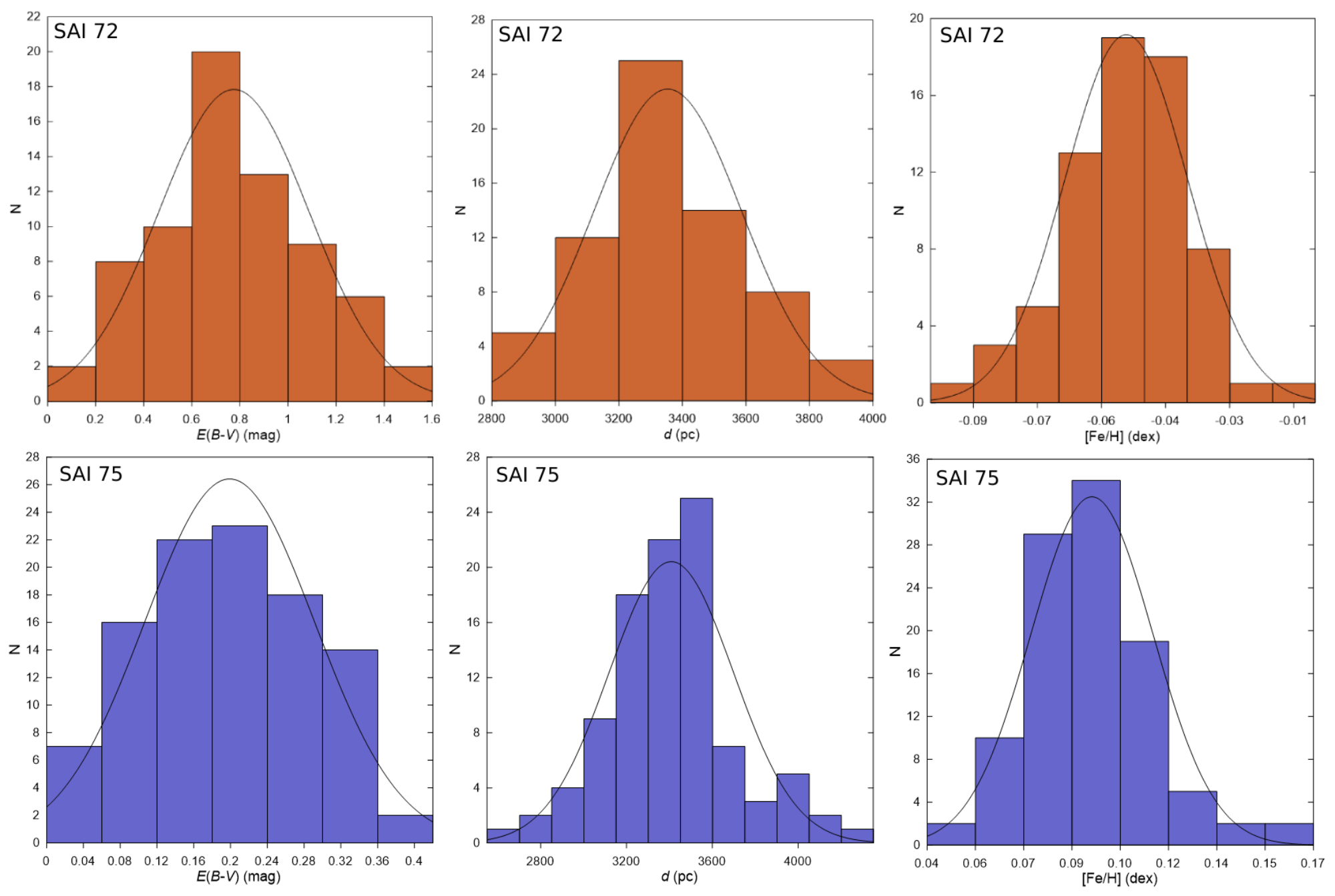}
\caption{The histograms display the distribution of color excess $E(B-V)$, distance ($d$), and metallicity [Fe/H] values for the members of SAI\,72 and SAI\,75, as determined from the SED analysis. The black lines overlaid on the distributions represent the corresponding standard Gaussian fits.}
\label{Fig: SED}
\end{figure*}

The histograms illustrating the \text{color} excess $E(B-V)$, distance $d$, and metallicity $[Fe/H]$ parameters derived from the SED analysis of member stars within SAI\,72 and SAI\,75 OCs are presented in Figure \ref{Fig: SED}. When the histograms of $E(B-V)$, $d$, and $[Fe/H]$ for all stars in the sample are examined, their distributions closely follow a Gaussian profile (see Figure \ref{Fig: SED}). By applying a Gaussian fit to the distance distribution, the mean distances ($d$) of the clusters are determined as $ 3353 \pm 233 $ pc for SAI\,72 and $ 3407 \pm 290 $ pc for SAI\,75. Similarly, the extinction histogram follows a Gaussian pattern, mean \text{color} excess $E(B-V)$ of $0.77 \pm 0.31 $ mag for SAI\,72 and $0.20 \pm 0.09$ mag for SAI\,75. The metallicity histogram also exhibits a Gaussian distribution, with the mean $[Fe/H]$ values calculated as $-0.05 \pm 0.01$ dex for SAI\,72 and $0.10 \pm 0.02$  dex for SAI\,75. 

To determine the optimal isochrone fit and extract astrophysical parameters, the metallicity values derived for SAI\,72 and SAI\,75 from the SED analysis were transformed into mass fraction $z$ \citep{Yontan2022, Elsanhoury2025,
Haroon2025}. This conversion follows the method provided by Bovy\footnote{https://github.com/jobovy/isodist/}, which is compatible with {\sc PARSEC} isochrone models \citep{Bressan2012}. The transformation is performed using the following equations:
\begin{equation}
z_{\rm x} = 10^{{\rm [Fe/H]} + \log \left( \frac{z_{\odot}}{1 - 0.248 - 2.78 \times z_{\odot}} \right)}
\end{equation}
\begin{equation}
z = \frac{(z_{\rm x} - 0.2485 \times z_{\rm x})}{(2.78 \times z_{\rm x} + 1)}.
\end{equation}
where, $z_{\rm x}$ and $z_{\odot}$ are intermediate parameters, with the adopted solar metallicity $z_{\odot}$ set at 0.0152 \citep{Bressan2012}. Using these equations, we derived the mass fraction values corresponding to the metallicities obtained from the SED analysis of SAI\,72 and SAI\,75: for SAI\,72, with [Fe/H] = -0.05 dex, the derived mass fraction is $z = 0.01363$. Similarly, for SAI\,75, with [Fe/H] = 0.10 dex, we determined the mass fraction as $z = 0.01886$. 

\subsubsection{Luminosity and Mass Function \\}
\begin{figure}
\centering
\includegraphics[width=0.9\linewidth]{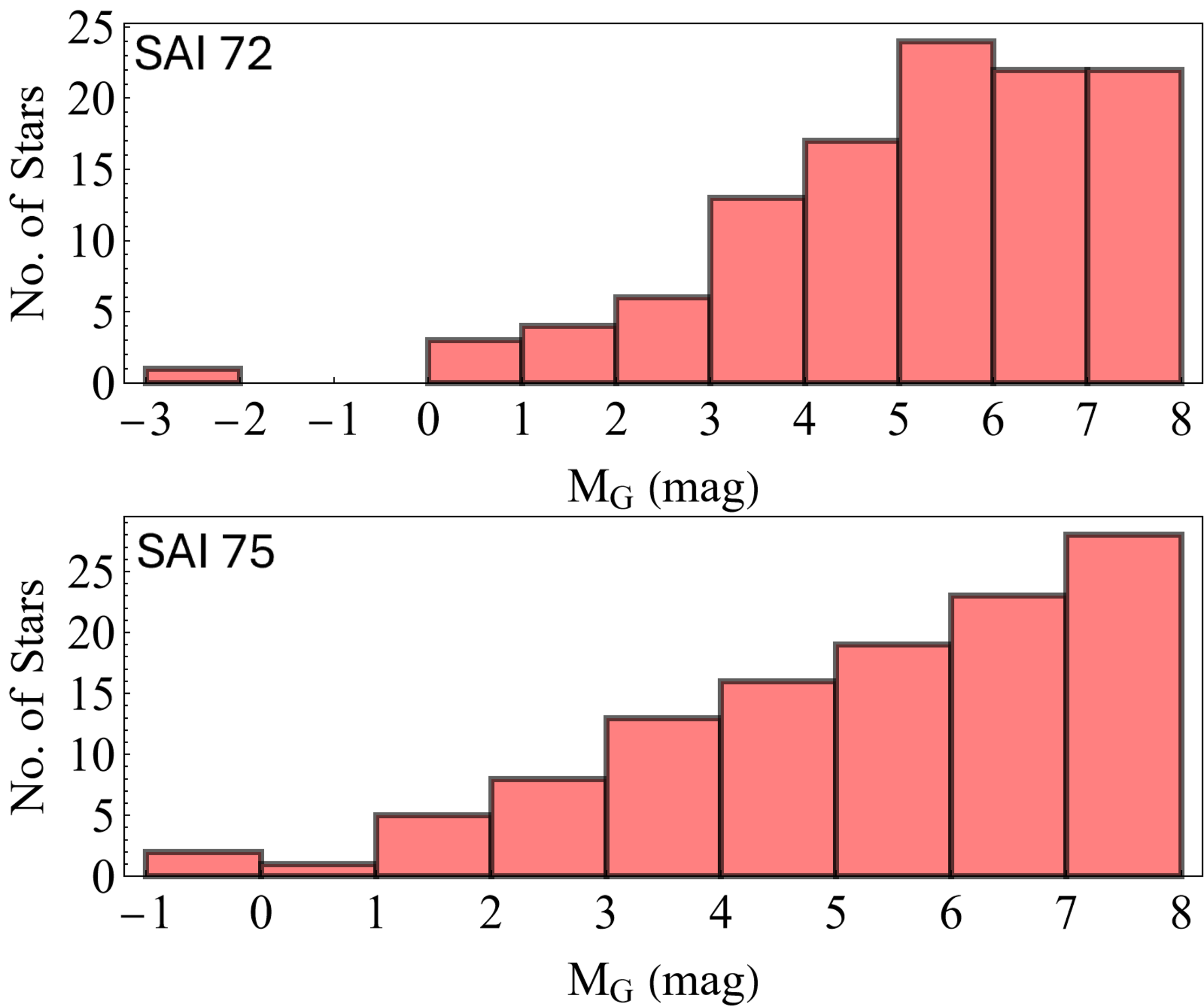}
\caption{The frequency distributions of the luminosity functions (LFs) in the $G$ band for the OCs under investigation.}
\label{Fig: 6}
\end{figure}

The luminosity function (LF) provides a statistical representation of the distribution of stars within a cluster across different absolute magnitude intervals \citep{Haroon2017}. To derive the LF for each cluster, we identified probable members and converted their apparent magnitudes into absolute magnitudes using the distance modulus. The resulting absolute magnitudes were then grouped into histograms with optimally chosen bin sizes to ensure a statistically significant star count, as illustrated in Figure \ref{Fig: 6}. Our analysis yields mean absolute magnitudes ($M_\text{G}$) of approximately 5.20 mag for SAI\,72 and 5.30 mag for SAI\,75.

\begin{figure}
\centering
\includegraphics[width=0.95\linewidth]{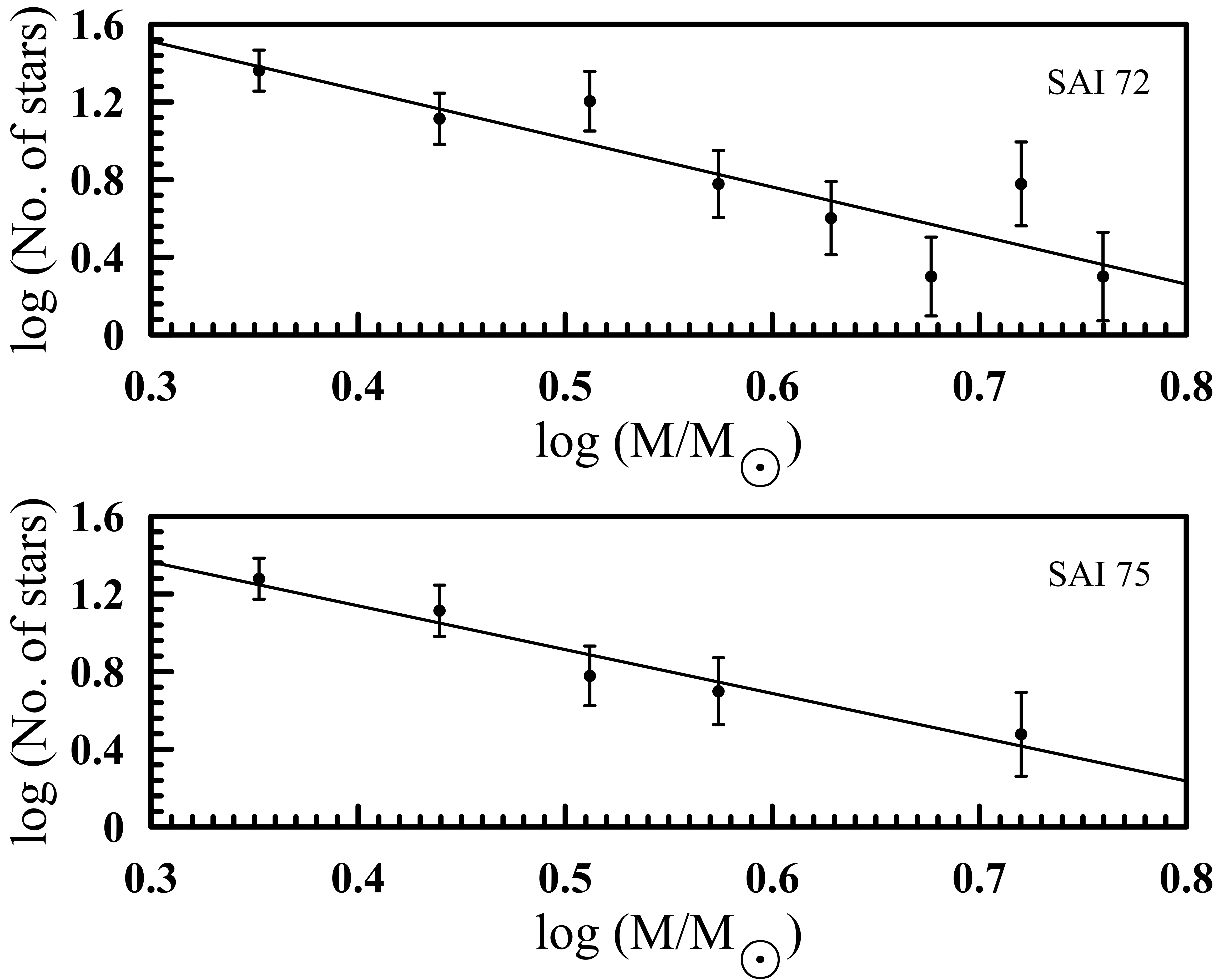}
\caption{The mass functions (MFs) for the clusters, fitted with the Salpeter power-law, are presented in the plots.}
\label{Fig: 7}
\end{figure}

The initial mass function (IMF) describes the mass distribution of stars at the time of their formation and is a key outcome of star formation processes \citep{Bisht2019}. By utilizing the LF and the corresponding mass function (MF), the IMF can be determined through the mass-luminosity relation (MLR). A seminal IMF formulation for stars in the solar neighborhood was introduced by \cite{Salpeter1955}, where the MF follows a power-law form:
\begin{equation} \label{Eq: 4} \dfrac{dN}{dM} \propto M^{-\alpha}. \end{equation}
where, $dN/dM$ denotes the number of stars within a given mass interval $(M, M + dM)$, and the exponent $\alpha$ is typically assigned a value of 2.35, implying that lower-mass stars dominate over their higher-mass counterparts. To estimate stellar masses within the clusters, we employed a polynomial mass-luminosity relation based on theoretical isochrone fitting:
\begin{equation} M_{\rm c}= a_0 + a_1 M_{\rm G} + a_{2} M_{\rm G}^2 + a_{3} M_{\rm G}^3 + a_{4} M_{\rm G}^4, \label{Eq: 5}
\end{equation}
where $a_0, a_1, a_2, a_3$, and $a_4$ are coefficients derived from the best-fit isochrones that describe the mass function of each cluster. Table \ref{Tab: 5} summarizes key parameters such as the mean absolute magnitudes ($M_\text{G}$), total cluster mass ($M_{\text{C}}$), mean stellar mass ($\overline{M_C}$), and the power-law slope ($\alpha$) obtained from our mass function analysis. The findings highlight variations in total stellar mass, member count, and mean mass per star, reflecting the clusters' differing evolutionary stages. The total masses were determined as $612\pm 174$ $M_{\odot}$ for SAI\,72 and $465\pm 90$ $M_{\odot}$ for SAI\,75. The mean stellar masses for these clusters are estimated to be $5.46\pm1.55$ $M_{\odot}$ and $4.04\pm0.78$ $M_{\odot}$, respectively. Figure \ref{Fig: 7} illustrates the mass functions of the clusters in logarithmic form.

\begin{table}[t]%
\centering
\caption{LF and MF for the clusters in the current study are presented.}%
\begin{tabular}{lcc}
\hline
\textbf{Parameters} & \textbf{SAI\,72}  & \textbf{SAI\,75}   \\
\hline
\hline
 $\overline{M_G} $ (mag) & 5.20   & 5.30     \\
 $M_C~(M_{\odot})$ & $612\pm 174$ & $465\pm 90$ \\
 $\overline{M_C}$ & $5.46\pm1.55$   & $4.04\pm0.78$   \\
$\alpha$ & 2.50 $\pm$ 0.02  & 2.26 $\pm$ 0.01 \\
\hline
\end{tabular}
\label{Tab: 5}
\end{table}

\subsection{Dynamical and Kinematical Properties}

\subsubsection{Dynamical Relaxation Timescale \\}
The internal dynamical evolution of an OC is primarily governed by gravitational interactions among its member stars. These interactions result in various dynamical processes, including mass segregation, stellar encounters, and energy redistribution. The characteristic timescale over which these processes drive the system toward equilibrium is known as the dynamical relaxation time \citep{Lamers2006}. This parameter is crucial in assessing whether a cluster retains its initial structure or has evolved into a dynamically relaxed state. The relaxation time is expressed as \citep{Spitzer1971}:
\begin{equation} \label{Eq. 6}
T_\text{relax} = \frac{8.9\times 10 ^5 N^{\frac{1}{2}} R^{\frac{3}{2}}_\text{h}}{(\overline{M_C})^{\frac{1}{2}} \log(0.4N)},
\end{equation}
where $N$ represents the number of stars in the cluster, $\overline{M}_{\rm C}$ represents the mean stellar mass (in $M_\odot$), and $R_{\rm h}$ (in pc) is the radius enclosing approximately 50$\%$ of the cluster mass. The latter can be derived using the relation \citep{Sableviciute_2006}:
\begin{equation}
R_{\rm h} = 0.547 \times r_{\rm c} \times \Big(\frac{r_{\rm t}}{r_{\rm c}}\Big)^{0.486},
\end{equation}
accordingly, $R_\text{h}$ are about 1.29 $\pm$ 0.01 and 1.17 $\pm$ 0.02 in parsecs for both SAI 72 and SAI 75, respectively. Therefore, we derived $T_{relax}$ of about 3.58 Myr and 3.61 Myr, respectively, indicating that both clusters have reached a dynamically relaxed state.

\subsubsection{Cluster Motion and Convergent Point \\}

\begin{figure}
\centering
\includegraphics[width=0.75\linewidth]{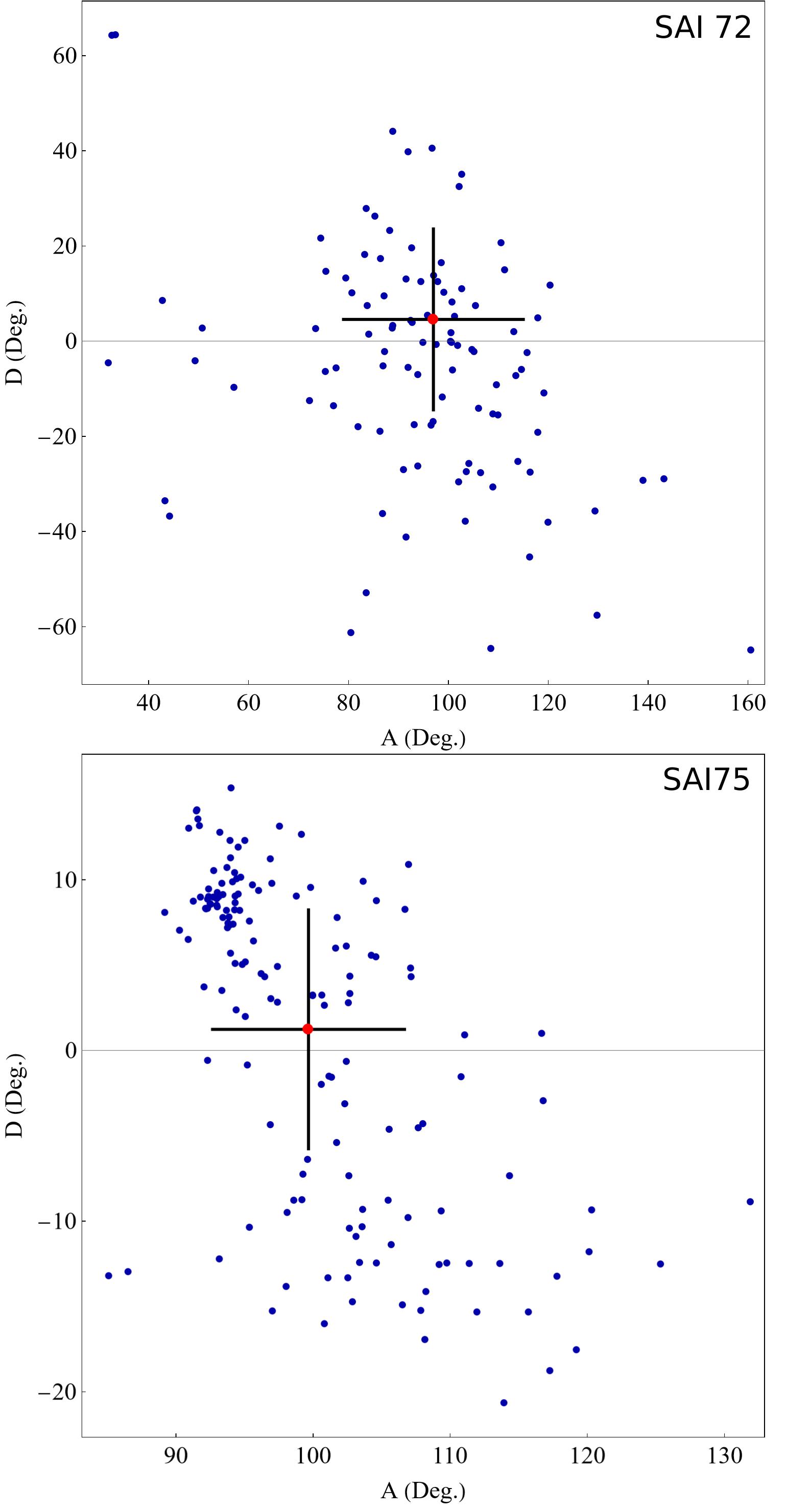}
\caption{The $AD$-diagram plots for the clusters, with the cross mark indicating the location of the apex point $(A,~D)_o$.}
\label{Fig: 8}
\end{figure}

The movement of an open cluster across the sky can be analyzed through its convergent point (CP), which marks the apparent direction of stellar motion in the celestial sphere \citep{Elsanhoury2018}. This point can be determined using different approaches, such as the classical CP method based on proper motion vectors or the $AD$ diagram technique, which incorporates radial velocities and astrometric parameters. In our study, we employed the $AD$ diagram method as described in \cite{Chupina2001,Chupina2006}.

\begin{figure*}
\centering
\includegraphics[width=0.99\linewidth]{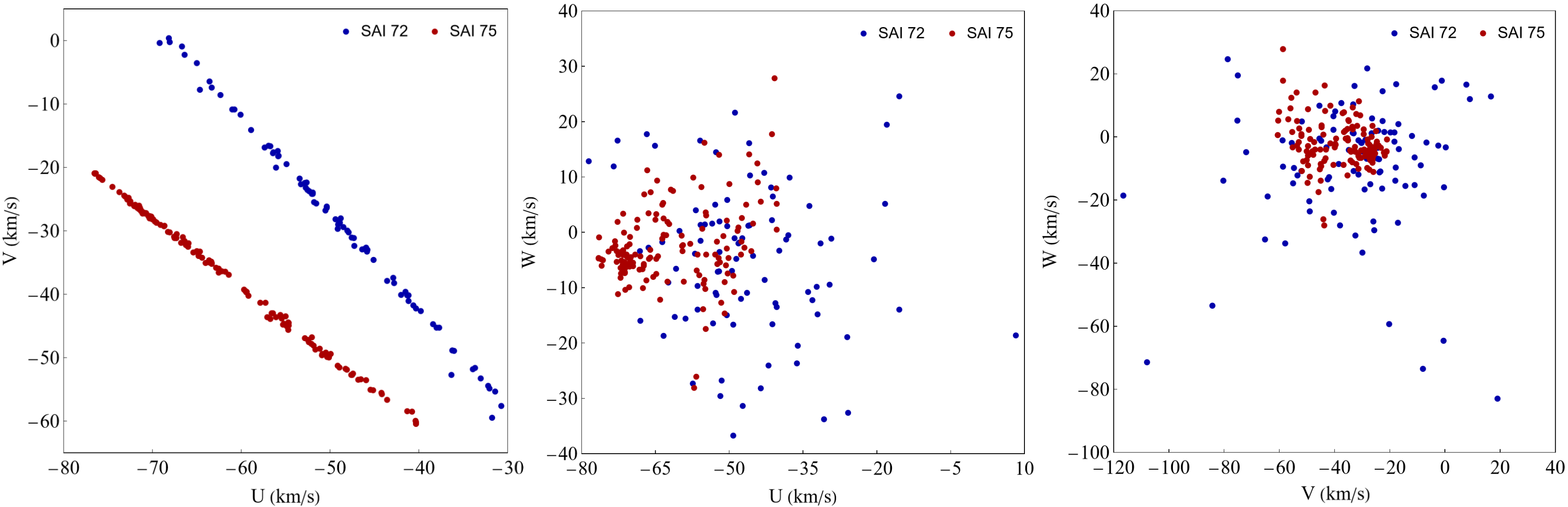}
\caption{The dispersion of the spatial velocity components along the Galactic coordinates of the clusters.}
    \label{Fig: 9}
\end{figure*}

The $AD$ diagram provides a visual representation of individual stellar apex coordinates within the cluster, derived from their spatial velocity components \citep{Maurya2021}. The equatorial coordinates of the convergent point $(A,~D)_o$ can be computed using:
\begin{equation}
\begin{split}
A_{\rm o} &= \tan^{-1} \Bigg(\frac{\overline{V_y}}{\overline{V_x}}\Bigg) \\
D_{\rm o} &= \tan^{-1} \Bigg(\frac{\overline{V_z}}{\sqrt {\overline{V_x^2} + \overline{V_{\rm y}^{2}}}}\Bigg)
\end{split}
\label{Eq. 7-8}
\end{equation}
where $V_{\rm x}, V_{\rm y},$ and $V_{\rm z}$ correspond to the space velocity components.

For cluster members at equatorial coordinates $(\alpha,~\delta)$ with distance $d$, proper motion $(\mu_{\alpha}\cos \delta,~\mu_{\delta})$, and radial velocity $V_{\rm r}$, the velocity components in the Cartesian frame are given by \cite{Melchior1958}:
\begin{equation} \label{Eq. 9}
\begin{split}
V_x &= -4.74~d~\mu_{\alpha}\cos{\delta}\sin{\alpha} - 4.74~d~\mu_{\delta}\sin{\delta}\cos{\alpha} \\
&+ V_{\rm r} \cos\delta \cos\alpha,
\end{split}
\end{equation}
\begin{equation} \label{Eq. 10}
\begin{split}
V_y &= +4.74~d~\mu_{\alpha}\cos{\delta}\sin{\alpha} - 4.74~d~\mu_{\delta}\sin{\delta}\cos{\alpha} \\
&+ V_{\rm r} \cos\delta \cos\alpha,
\end{split}
\end{equation}
\begin{equation} \label{Eq. 11}
\begin{split}
V_{\rm z} &= +4.74~d~\mu_\delta\cos\delta + V_{\rm r}\sin\delta.
\end{split}
\end{equation}
Following the determination of mean velocity components $V_{\rm x}, V_{\rm y},$ and $V_{\rm z}$, the cluster's apex coordinates were computed using the $AD$ method, yielding $(A,~D)_o = (^{o}.016 \pm 0^{o}.09,~4^{o}.573 \pm 0^{o}.05)$ for SAI\,72 and $(99^{o}.677 \pm 0^{o}.10,~1^{o}.243 \pm 0^{o}.09)$ for SAI\,75. Figure \ref{Fig: 8} illustrates the convergent point locations for both clusters, while their respective coordinate values are detailed in Table \ref{Tab: 6}. The conversion from equatorial to Galactic velocity components follows the transformation equations:
\begin{equation} \label{Eq: 12}
\begin{split}
U &= -0.0518807421 V_{\rm x} -0.872222642 V_{\rm y} \\
&- 0.48634200 V_{z},
\end{split}
\end{equation}
\begin{equation} \label{Eq: 13}
\begin{split}
V &= +0.4846922369 V_{\rm x} -0.4477920852 V_{\rm y} \\
&+ 0.7513692061 V_{z},
\end{split}
\end{equation}
\begin{equation} \label{Eq: 14}
\begin{split}
W &= -0.873144899 V_{\rm x} -0.196748341 V_{\rm y} \\
&+ 0.4459913295 V_{\rm z}.
\end{split}
\end{equation}
The mean space velocities $(\overline{U},~\overline{V},~\overline{W})$ were computed using:
\begin{equation} \label{eq: 15}
\overline{U} = \dfrac{1}{N}\sum ^{N}_{i=1}U_i,\quad
\overline{V} = \dfrac{1}{N}\sum ^{N}_{i=1}V_i, \quad
\overline{W} = \dfrac{1}{N}\sum ^{N}_{i=1}W_i.
\end{equation}
Figure \ref{Fig: 9} presents the velocity distribution of member stars in Galactic coordinates, with the computed mean values summarized in Table \ref{Tab: 6}.

\subsubsection{Elements of Solar Motion \\} 
The determination of the solar motion parameters, including the solar apex coordinates ($l_{\rm A},~b_{\rm A}$) and the velocity magnitude ($S_{\odot}$), has been the subject of numerous studies involving various stellar populations and interstellar tracers within the Milky Way. Observational techniques used for this purpose generally fall into three main categories: those based solely on radial velocities, those relying exclusively on proper motions, and hybrid methods incorporating both radial and transverse velocity components \cite{Fouad2003}.

In this study, we determine the solar velocity components ($U_{\odot},~V_{\odot},~W_{\odot}$) by analyzing the motion of a sample of nearby stars. This method treats the selected stars as a reference frame to infer the Sun's movement relative to the surrounding stellar system. Given the mean velocity components ($\overline{U},~\overline{V},~\overline{W}$) of a stellar group in the Galactic coordinate system, the corresponding solar motion components in km s$^{-1}$ are calculated using the following relations \citep{Elsanhoury2016,Elsanhoury2022}:
\begin{equation} \label{Eq.16}
U_{\odot}= -\overline{U},\; V_{\odot}= -\overline{V}, \; \text{and} \; W_{\odot}= -\overline{W}.
\end{equation}
From these components, the magnitude of the solar velocity relative to the selected stars can be derived as:
\begin{equation} \label{Eq. 17}
S_{\odot}=\sqrt{(\overline{U})^2+(\overline{V})^2+(\overline{W})^2}.
\end{equation}
Furthermore, the solar apex position in Galactic coordinates ($l_A,~b_A$) is determined using:
\begin{equation} \label{Eq. 18}
l_\text{A} = \tan^{-1}\left(\frac{-\overline{V}}{\overline{U}}\right)\;\; \text{and} \;\;
b_\text{A} = \sin^{-1} \left(\frac{-\overline{W}}{S_\odot}\right),  
\end{equation}
where $l_\text{A}$ and $b_\text{A}$ denote the Galactic longitude and latitude of the solar apex, respectively. The computed solar velocity magnitude and apex coordinates for each analyzed stellar group are presented in Table \ref{Tab: 6}.

\begin{table}
\centering
\renewcommand{\arraystretch}{1.4}
\setlength{\tabcolsep}{2.8pt}
\caption{The dynamical and kinematic parameters for the SAI\,72 and SAI\,75 clusters are presented.}
\begin{tabular}{l|ccc}
\hline
\textbf{Parameter} & \textbf{SAI\,72} & \textbf{SAI\,75} \\
\hline

& \multicolumn{3}{c}{Kinematical Parameters} \\
\hline
\hline
$T_{\rm relax}$ (Myr) & 3.58 & 3.61\\
$\overline{V_x}$ (km s$^{-1}$) & ~$-$6.96$\pm$0.38 & $-$12.14$\pm$3.48\\
$\overline{V_y}$ (km s$^{-1}$) & 56.56$\pm$7.52& ~71.17$\pm$8.44\\
$\overline{V_z}$ (km s$^{-1}$) & ~~4.56$\pm$0.47& ~~~1.57$\pm$0.08\\
$\overline{U}$ (km s$^{-1}$)& $-$51.19$\pm$7.15& $-$62.21$\pm$7.90 \\
$\overline{V}$ (km s$^{-1}$)& $-$5.28$\pm$0.44 & $-$36.58$\pm$6.05\\
$\overline{W}$ (km s$^{-1}$)& $-$3.02$\pm$0.06 & ~$-$2.71$\pm$0.06\\
$S_{\odot}$ (km s$^{-1}$)&  ~58.42$\pm$7.64  &~72.22$\pm$8.50 \\
\hline
& \multicolumn{3}{c}{Dynamical Parameters} \\
\hline
\hline
$A~(^o)$& 97.016 $\pm$ 0.09  & 99.677 $\pm$0.10 \\
$D~(^o)$& 4.573 $\pm$ 0.05 & 1.243 $\pm$ 0.09\\ 
$l_A~(^o)$ &$-$30.56&$-$30.452\\
$b_A~(^o)$ &12.88&2.149\\
$Z_{\rm max}$ (kpc)& 0.109$\pm$0.009 & ~0.232$\pm$0.024  \\
$R_{\rm a}$ (kpc)& 11.156$\pm$0.050  & 10.992$\pm$0.143  \\
$R_{\rm p}$ (kpc)& 10.692$\pm$0.045  & ~6.743$\pm$0.1  \\
$R_{\rm m}$ (kpc)& 10.924$\pm$0.047 &  ~8.867$\pm$0.170 \\
$e$              &  0.02$\pm$0.00&0.24$\pm$0.02 \\
$T_{\rm p}$ (Myr)&  316$\pm$2.00 & 253$\pm$1.00 \\
$R_{\rm Birth}$ (kpc)& \text{10.825$\pm$0.068} & \text{9.583$\pm$0.231}\\
\hline
\end{tabular}
\label{Tab: 6}
\end{table}

\subsubsection{Dynamic Orbit Parameters \\}

\begin{figure*}
\centering
\includegraphics[width=0.9\linewidth]{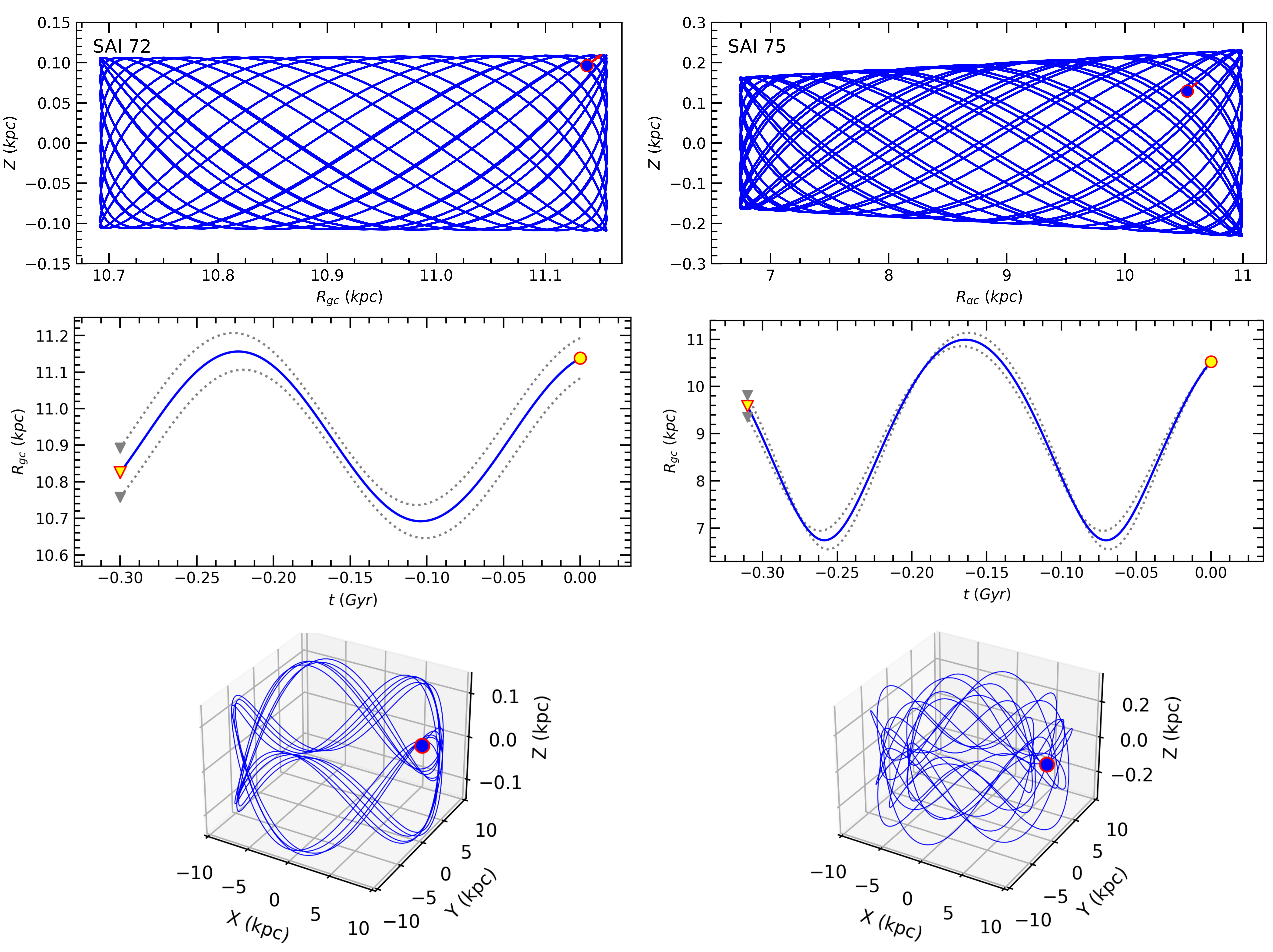}
\caption{The Galactic orbits and birth radii of SAI\,72 and SAI\,75 are shown in three different planes: $Z~\times~R_{\rm gc}$ (top), $R_{\rm gc}$ $\times$ $t$ (middle), and $X$ $\times$ $Y$ $\times$ $Z$ (bottom). In the middle, present-day positions are shown by filled yellow circles, and birth positions are marked by filled triangles. The gray dotted line represents the orbit considering input parameter errors. The filled triangles show the lower and upper error estimates of the birth locations.}
\label{Fig: Orbit}
\end{figure*}

The methodology adopted in this study has been successfully applied to both individual stars and OCs, enabling a comprehensive examination of their spatial distribution and kinematic characteristics in the Milky Way \citep{YontanCanbay2023, Tasdemir2023, Tasdemir2025b}. This approach is essential for investigating the structural evolution of our Galaxy. The Galactic properties of SAI\,72 and SAI\,75 were analyzed by assessing their kinematic and orbital parameters \citep{Dursun2024, Tasdemir2025a}. Specifically, the space velocity components, Galactic orbital elements, and formation radii of these clusters were determined. For orbital calculations, we utilized the {\sc MWPotential2014} model, implemented within the {\sc galpy}\footnote{https://galpy.readthedocs.io/en/v1.5.0/} package, a widely used tool in Galactic dynamics developed by \citet{Bovy2015}. This model incorporates key parameters such as the Galactocentric distance ($R_{\rm gc} = 8.20 \pm 0.10$ kpc) and the Sun’s orbital velocity ($V_{\rm rot} = 220$ km s$^{-1}$) \citep{Bovy2012, Bovy2015}, with the vertical displacement of the Sun from the Galactic plane taken as $Z_{\rm 0} = 25 \pm 5$ pc \citep{Juric2008}.
The input data for the analysis include the equatorial coordinates ($\alpha$,~$\delta$), distance ($d$), and the mean proper motion components ($\mu_\alpha\cos\delta,~\mu_\delta$) from Table \ref{Tab: 4}. The present-day positions of SAI 72 and SAI 75 were estimated by integrating their orbits forward in time with 1 Myr steps to their respective ages. The apogalactic and perigalactic distances (\(R_{\rm a}\) and \(R_{\rm p}\)), the mean Galactocentric distance (\(R_{\rm m}=(R_{\rm a}+R_{\rm p})/2\)), the eccentricity (\(e\)), and the maximum vertical distance from the Galactic plane (\(Z_{\rm max}\)) were determined for each cluster. The computed parameters for SAI\,72 and SAI\,75 are summarized in Table \ref{Tab: 6}. The maximum height above the Galactic plane (\(Z_{\rm max}\)) was determined as \(109 \pm 9\) pc for SAI\,72 and \(232 \pm 24\) pc for SAI\,75, supporting their classification as members of the Milky Way’s young stellar disk \citep{Guctekin2019}. These findings are depicted in Figure \ref{Fig: Orbit} (top). The three-dimensional orbital trajectories of both clusters, illustrated in Figure \ref{Fig: Orbit} (bottom), reveal near-circular orbits with gradual oscillations relative to the Galactic plane. Furthermore, the dynamical analysis suggests that both SAI\,72 and SAI\,75 originated beyond the solar circle. Specifically, SAI\,72 has a birth radius of \( R_{\rm Birth} = 10.824 \pm 0.068 \) kpc, while SAI\,75 has a birth radius of \( R_{\rm Birth} = 9.583 \pm 0.231 \) kpc, as shown in Figure \ref{Fig: Orbit} (middle).

\section{CONCLUSION}\label{sec4}
We performed an extensive photometric and kinematic study of the open clusters SAI\,72 and SAI\,75 using data from $Gaia$DR3. By determining stellar membership probabilities, we identified 112 stars in SAI\,72 and 115 stars in SAI\,75 with probabilities exceeding 50\%. The structural properties of these clusters were examined through a combination of photometric and kinematic methodologies. Our analysis was conducted using the \text{ASteCA} code, yielding the following main results:

The recalculated central coordinates of the clusters are $\alpha = 103^{\circ}.958$ ($6^{\text{h}}55^{\text{m}}50.39^{\text{s}}$), $\delta = 0^{\circ}.217$ ($+00^{\circ}12'01.58''$) for SAI\,72 and $\alpha = 108^{\circ}.836$ ($7^{\text{h}}15^{\text{m}}18.27^{\text{s}}$), $\delta = -7^{\circ}.429$ ($-7^{\circ}25'44.4''$) for SAI\,75. These values align with previous findings from \cite{Glushkova2010}. The radial density profiles (RDP) provided cluster size estimates, yielding limiting radii of 2.35 and 2.19 arcminutes for SAI\,72 and SAI\,75, respectively. The core radii, where the stellar density reaches half of its peak value, were computed as $r_{\rm c} = 1.93$ and $1.56$ arcminutes.

Using the most probable members, we constructed color-magnitude diagrams (CMDs) based on $Gaia$DR3 photometric data, estimating the ages of the clusters to be 316 Myr for SAI\,72 and 302 Myr for SAI\,75. The corresponding distances were derived as $3160 \pm 80$ pc and $3200 \pm 200$ pc for SAI\,72 and SAI\,75, respectively. We also examined their spatial distribution in the Galaxy, determining their galactocentric distances and positions relative to the Galactic plane $(X_\odot,~Y_\odot,~Z_\odot)$.

The SED-derived parameters for both clusters were systematically compared with the results obtained from isochrone fitting to assess their consistency and reliability. This comparative analysis aimed to minimize the degeneracies inherent in the isochrone-fitting method, particularly in the estimation of age ($t$), metallicity [Fe/H], and color excess $E(B-V)$. By incorporating SED constraints, which leverage multi-wavelength photometric data and spectral energy distribution modeling, a more refined determination of cluster properties was achieved. The combined approach provided a more precise astrophysical characterization by addressing degeneracies in the isochrone fitting method, which often result in uncertainties in age, metallicity, and reddening estimates. By using an additional independent SED fitting method to independently test these parameters, we were able to reduce these degeneracies and improve the precision and accuracy of the fundamental parameter estimates.

To analyze the mass and luminosity distribution, we derived the luminosity functions (LFs) and mass functions (MFs) from the identified cluster members. The total cluster masses were estimated as $612 \pm 174~M_\odot$ for SAI\,72 and $465 \pm 90~M_\odot$ for SAI\,75, with corresponding MF slopes of -2.50 $\pm$ 0.02 and -2.26 $\pm$ 0.01. The dynamical study indicates that both clusters have attained a state of near dynamical equilibrium.

The space velocity components allowed us to determine the apex coordinates using the $AD$ diagram method. The calculated apex positions were $(A,~D)_o = (97^{o}.016~\pm~0^{o}.09,~4^{o}.573~\pm~0^{o}.05)$ for SAI\,72 and $(99^{o}.677 \pm 0^{o}.10,~1^{o}.243~\pm~0^{o}.09)$ for SAI\,75. Furthermore, we applied transformation equations to convert equatorial velocity components into the Galactic frame, deriving the space velocity components $(U,~V,~W)$. The solar motion parameters for both clusters were also estimated, providing insights into their kinematic properties relative to the Galactic system.

By examining the Galactic properties of SAI\,72 and SAI\,75, including their kinematic and orbital parameters, we determined crucial aspects such as space velocity components, Galactic orbital elements, and formation radii. The eccentricity of the orbit for SAI\,72 is \( e = 0.02\pm0.00 \), which indicates a nearly circular orbit. This low eccentricity suggests that the cluster's orbit is highly stable, with minimal variations in its distance from the Galactic center. Such a circular orbit implies that SAI\,72 has remained relatively unaffected by significant gravitational perturbations over time, allowing it to maintain a consistent path within the Galactic disk. In contrast, SAI\,75 has a higher eccentricity of \( e = 0.24\pm0.02 \), indicating a more elongated orbit with greater variations in its position relative to the Galactic center. This difference in orbital characteristics may reflect distinct dynamical histories and environmental interactions for the two clusters. The maximum vertical distances from the Galactic plane were found to be 109 $\pm$ 9 pc for SAI\,72 and 232 $\pm$ 24 pc for SAI\,75, affirming their classification as members of the Milky Way’s young stellar disk. Furthermore, the dynamical analysis indicated that SAI\,72 formed within the solar circle, with a birth radius of $10.824 \pm 0.068$ kpc, while SAI\,75 originated beyond the solar circle at a birth radius of $ 9.583 \pm 0.231$ kpc.

\section*{Acknowledgements}
We sincerely thank the anonymous referee for their valuable suggestions, which greatly enhanced the quality of this paper. This work presents results from the European Space Agency (ESA) space mission $Gaia$. $Gaia$ data are being processed by the $Gaia$ Data Processing and Analysis Consortium (DPAC). Funding for the DPAC is provided by national institutions, in particular, the institutions participating in the $Gaia$ Multi-Lateral Agreement (MLA). The $Gaia$ mission website is \url{https://www.cosmos.esa.int/gaia}. The $Gaia$ archive website is \url{https://archives.esac.esa.int/gaia}.  The authors would like to express their gratitude to the Deanship of Scientific Research at Northern Border University, Arar, KSA, for funding this research under project number "NBU-FFR-2025-237-06".

\bibliography{JAA}

\end{document}